\documentclass[acmsmall]{acmart}
\AtBeginDocument{%
  }

\setcopyright{acmcopyright}
\copyrightyear{2025}
\acmYear{2025} \acmMonth{4}
\acmDOI{XXXXXXX.XXXXXXX}
\thanks{© 2025 ACM. This is the author’s version of the work that has been accepted for publication in Proceedings of the ACM on Human-Computer Interaction (PACMHCI), CSCW 2025. Personal use of this material is permitted. Permission from ACM must be obtained for all other uses, including but not limited to reprinting/republishing this material for advertising or promotional purposes, creating new collective works, for resale or redistribution to servers or lists, or reuse of any copyrighted component of this work in other works. The final version will be available at: \url{https://doi.org/10.1145/3710987}.}

\begin{document}

\title{From Interaction to Attitude: Exploring the Impact of Human-AI Cooperation on Mental Illness Stigma}

\author{Tianqi Song}
\email{tianqi_song@u.nus.edu}
\orcid{0000-0001-6902-5503}
\affiliation{%
  \institution{National University of Singapore}
  \country{Singapore}
}

\author{Jack Jamieson}
\affiliation{%
  \institution{NTT}
  \country{Japan}}
\email{jack@jackjamieson.net}
\orcid{0000-0002-8444-5722}

\author{Tianwen Zhu}
\affiliation{%
  \institution{National University of Singapore}
  \country{Singapore}}
\email{tianwen@g.ucla.edu}
\orcid{0009-0008-0594-1084}

\author{Naomi Yamashita}
\affiliation{%
  \institution{NTT}
  \country{Japan}}
\email{naomiy@acm.org}
\orcid{0000-0003-0643-6262}

\author{Yi-Chieh Lee}
\affiliation{%
  \institution{National University of Singapore}
  \country{Singapore}}
\email{yclee@nus.edu.sg}
\orcid{0000-0002-5484-6066}

\renewcommand{\shortauthors}{Tianqi Song et al.}

\begin{abstract}
  AI conversational agents have demonstrated efficacy in social contact interventions for stigma reduction at a low cost. However, the underlying mechanisms of how interaction designs contribute to these effects remain unclear.
This study investigates how participating in three human-chatbot interactions affects attitudes toward mental illness. 
We developed three chatbots capable of engaging in either one-way information dissemination from chatbot to a human or two-way cooperation where the chatbot and a human exchange thoughts and work together on a cooperation task.
We then conducted a two-week mixed-methods study to investigate variations over time and across different group memberships. 
The results indicate that human-AI cooperation can effectively reduce stigma toward individuals with mental illness by fostering relationships between humans and AI through social contact. Additionally, compared to a one-way chatbot, interacting with a cooperative chatbot led participants to perceive it as more competent and likable, promoting greater empathy during the conversation. However, despite the success in reducing stigma, inconsistencies between the chatbot’s role and the mental health context raised concerns.
We discuss the implications of our findings for human-chatbot interaction designs aimed at changing human attitudes.
\end{abstract}



\received{January 2024}
\received[revised]{July 2024}
\received[accepted]{October 2024}

\maketitle

\section{Introduction}

Stigma is the social rejection of individuals with attributes that are deeply discredited within their societies \cite{goffman2009stigma}. It has lasting negative impacts on public health and social equality, particularly for individuals with mental illness \cite{corrigan2002understanding, goffman2009stigma}, who often delay accessing health services due to stigma and fear of discrimination \cite{corrigan2004stigma}. Stigmatizing thoughts about people with lived experience of mental illness – notably, that they are violent, dangerous, and dependent – also causes individuals to feel isolated at work and face discrimination there \cite{peluso2009public}. Consequently, mental illness stigma remains a pervasive social issue that profoundly affects the well-being of individuals worldwide \cite{levinson2010associations}.

Various strategies have been proposed to reduce mental illness stigma \cite{heijnders2006fight}. These have included advocacy, protest, education, and social contact \cite{corrigan1999lessons}. 
However, research has indicated that attempts to mitigate stereotypes through protest and education can result in a rebound effect \cite{corrigan1999lessons}. 
In contrast, there is substantial evidence that interventions based on social contact can effectively reduce stigma \cite{makhmud2022indirect}. These interventions can also be integrated into clinical practice to improve healthcare services for people with mental illnesses \cite{lien2021challenging, lanfredi2019effects, abi2019knowledge}.

With the advancement of technology, 
researchers have investigated digital social contact interventions to reduce social stigma, such as e-contact, virtual reality animation, and virtual agents \cite{figoli2022ai, hobert2020small, maestre2023s, kim2020helping, lee2023exploring}. 
Among these technologies, chatbots have proven effective in reducing stigmatizing attitudes \cite{lee2023exploring} while requiring minimal human involvement and effort. For example, research \cite{lee2023exploring} has shown that introducing chatbots as virtual agents with mental illnesses that engage in conversations with humans, is an effective strategy for reducing stigma.

Although interactions between humans and chatbots have shown effectiveness in reducing stigma \cite{lee2023exploring}, there is still insufficient research and understanding in this area. 
On the one hand, as chatbots gain proficiency in presenting designated and flexible conversations, there are more opportunities for them to act as social actors and exert influence on humans \cite{skjuve2022longitudinal, grove2021co, skjuvechatbots}. This underscores the importance of understanding how human-chatbot cooperation influences the transformation of human attitudes, especially when the chatbot represents a stigmatized identity.
On the other hand, previous studies focused solely on chatbots that deliver depressed stories to humans \cite{lee2019caring, lee2023exploring}, relying on script-based conversations without promoting diverse interactions. This simplistic design limits understanding of user perceptions towards the chatbot and interaction process, raising questions about whether users find the interaction positive or enjoyable and its potential effectiveness in real-world applications.

To understand and uncover effective social contact designs, we chose two key elements from previous in-person social contact studies to investigate their effects on stigma reduction and user experience: 1. "interaction mode" (cooperation vs. non-cooperation) \cite{ho2017reducing, kohrt2021collaboration, evans2012mass} and 2. "content topics" (mental-illness related vs. unrelated) \cite{lanfredi2019effects, rademaker2020applying, maunder2019intergroup}. Though these two elements have been discussed separately in previous work about human-human social contact, there is a lack of research on their combined effects. Moreover, these previous studies in reducing stigma generally employ survey or campaign methods without control groups, leading to issues with the generalizability of results \cite{jorm2020effect}. Furthermore, it is unknown whether conclusions drawn from human-human interactions can be applied to chatbots, as human-chatbot interactions are more limited than human-human interactions. Therefore, a comparative study that adapts these two design elements in a human-chatbot scenario is necessary.

Accordingly, the present study investigates whether and to what extent different interaction designs with chatbots can reduce mental illness stigma. Specifically, we constructed three chatbots that delivered vignettes about ‘their’ experiences of having depression. The first chatbot provided mental illness-related material for participants to read (an \textit{information dissemination task}). The second, using the same mental illness material, engaged in a human-chatbot \textit{cooperation task} with its user. The third shared scientific knowledge unrelated to depression while engaging in the same cooperation task as the second one. 
We divided the participants into three groups, each corresponding to one of these chatbots, and measured their attitudes toward people with mental illness before and after a two-week user study. 
Additionally, we conducted post-study interviews to gain insight into participants' interaction experiences, reflections during social interaction, and their potential willingness to accept the chatbot.
Our analysis of both quantitative and qualitative data revealed the impacts of human-chatbot interactions on participants' impressions and understanding of mental illness. This comprehensive examination contributes to the discourse on stigma reduction and informs future applications of virtual agent technology.

This work makes several contributions to the CSCW community.

\begin{itemize}
    \item First, our study addressed a gap in understanding how human-AI cooperation can foster attitude change toward stigmatized groups. This research expands the application of human-AI cooperation to social impact scenarios, contributing valuable insights into chatbot design and its role in shaping attitudes toward specific social groups.
    \item Second, we examined three interaction designs to reduce stigma toward individuals with mental illness. While all designs effectively decreased participants' overall stigma, inconsistencies between the chatbot’s mental health context and task topics led to unintended negative effects. These findings highlight the importance of aligning the AI agent’s role and messaging with its context to positively influence attitudes toward mental illness stigma.
    \item Additionally, our research provides empirical evidence on how varying human-AI interactions influence users' impressions and experiences during social contact. By showing that cooperative interactions foster more positive user impressions, we advocate for implementing cooperative human-AI designs in future applications, as these may promote greater user acceptance and engagement.
\end{itemize}
\section{Related Work}

\subsection{Mental illness Stigma Definitions and Interventions}

Stigmatization of people with mental illnesses is often rooted in societal stereotypes and prejudices but can also involve discriminatory behaviors that diverge from accepted norms \cite{corrigan2002understanding, goffman2009stigma}. 
It can have serious repercussions, such as making people unwilling to seek mental-health treatment due to fear of being stigmatized \cite{corrigan2004stigma}, perpetuating historical injustices against marginalized groups \cite{pendse2022treatment}, and limiting access to employment, housing, and social interaction \cite{corrigan2002understanding}. 
Past research has identified two types of stigma \cite{rusch2005mental}: "self-stigma," which involves internalized shame and negative attitudes individuals with mental illness may have about their own condition, and "social stigma" (or "public stigma"), which encompasses the negative or discriminatory attitudes others may hold about mental illness. Our study focuses on reducing social stigma.

The negative impact of the stigma of mental illness has attracted the attention of governments \cite{barry2014stigma}, WHO \footnote{https://www.emro.who.int/mnh/campaigns/anti-stigma-campaign.html}, social groups \footnote{https://www.irrsinnig-menschlich.de} \footnote{https://www.sane.org/} \footnote{https://www.sst.dk/da/en-af-os/ONE-OF-US} and non-governmental organizations (e.g., World Psychiatric Association) \cite{rusch2005mental}. 
To reduce stigma, three main strategies have been employed: protest, education, and social contact~\cite{corrigan1999lessons}.
The first usually involves challenging biased and stereotypical views of mental health problems \cite{corrigan2002understanding}. 
Protests can convey emotions, grab attention, and increase societal awareness \cite{rusch2005mental}. 
However, they may also foster negative perceptions about the protesting group \cite{monteith1998suppression, corrigan2001three}, rendering them less suitable for widespread implementation.
Education, meanwhile, provides information to help people form more reasoned opinions about mental illness \cite{corrigan1999lessons, corrigan2002understanding}.
While education leads to short-term enhancements in attitudes \cite{griffiths2014effectiveness, stubbs2014reducing, pinfold2003reducing}, the extent and duration of these improvements in attitudes and behavior may be restricted and weaker when compared to social-contact methods \cite{dobson2022myths}.
Among all these approaches, social contact interventions have been proven to be the most effective in reducing intergroup stigma \cite{maunder2019intergroup, thornicroft2016evidence}. 
Therefore, our study focuses on social contact interventions, comparing and evaluating multiple approaches.

\subsection{Social Contact Interventions}
\label{sec:stigma-intervention}

Social-contact interventions involve facilitating positive social interactions among individuals in the general population and those experiencing mental health problems \cite{corrigan2003attribution}.
In discussions of how such social contact might be optimized, the Intergroup Contact Hypothesis \cite{allport1954nature, brown2005integrative} stands out as a seminal framework for mitigating prejudice between members of majority and minority groups.  
Based on the Contact Hypothesis, Allport et al. \cite{allport1954nature} outlined four conditions that are conducive to reducing prejudice: 1) equal status, 2) shared goals, 3) intergroup cooperation, and 4) endorsement from authorities, laws, or customs.
Specifically, equal status implies that neither group holds a position of power or superiority over the other, and shared goals refer to members of different groups working together towards common objectives. 
Both these factors can promote positive inter-group relations and boost inter-group cooperation’s prejudice-reducing effects \cite{desforges1991effects, pettigrew2006meta, pettigrew2008does}.

Researchers have investigated different designs to enhance interaction between stigmatized groups and individuals who have stigmatizing attitudes, knowledge, or behavior \cite{adu2022social}, which could be categorized into three design types:

\textit{Design 1} involves sharing information about mental illness, including personal experiences, scientific knowledge, and recovery processes \cite{lee2023exploring, rodriguez2021controlled, stelzmann2021can, yuen2021effects, brown2020effectiveness}. For example, Rodriguez et al. \cite{rodriguez2021controlled} presented an online program on mental disorder recovery to stigmatized groups and university students.
\textit{Design 2}, inspired by Allport's theory, involves collaborative tasks focused on mental illness topics \cite{ho2017reducing, kohrt2021collaboration}. Kohrt \cite{kohrt2021collaboration}, for instance, had individuals take photos representing their experiences with mental illness. These photos were compiled into a video with voiceovers and text, followed by a discussion between the audience and the video creator.
\textit{Design 3} incorporates collaborative tasks on general topics unrelated to mental illness \cite{desforges1991effects, maunder2019modern}. For example, Desforges \cite{desforges1991effects} used cooperative learning, where participants read materials and taught each other to achieve a comprehensive understanding and positive evaluation.

Based on the three design approaches, we identify two critical design elements:
\textbf{1) interaction mode}, encompassing one-way and two-way interactions (e.g., "information dissemination" vs. "cooperation task"; and \textbf{2) content topics} (e.g., mental illness topics vs. other topics).
While the effects of these elements on stigma reduction have been discussed in prior literature, some conclusions have been inconsistent.
Regarding \textbf{interaction mode}: Allport's theory~\cite{allport1954nature} asserts that cooperation is a key factor for reducing stigma, and this has been been supported by research about anti-stigma campaigns~\cite{ho2017reducing, kohrt2021collaboration} and community surveys \cite{evans2012mass}. 
However, these studies have been criticized for having low methodological quality, particularly by lacking control groups, which weakens the generalizability of their conclusions \cite{jorm2020effect}. 
Regarding \textbf{content topics}, some studies suggest that mental-illness content aids stigma reduction by enhancing knowledge and fostering mutual understanding \cite{lanfredi2019effects, rademaker2020applying}. However, a meta-analysis argued that the presence or absence of mental-illness content in social contact has no differential impact on stigma reduction \cite{maunder2019intergroup}.

In general, the low generalizability and inconsistencies in these findings are primarily due to the absence of rigorously controlled experiments comparing interaction modes and content topics in social contact \cite{jorm2020effect}. This limits our understanding of whether cooperation or information alone can effectively reduce stigma and raises concerns about the potential misallocation of resources in designing cooperative interventions. Thus, a comparative study design is necessary.

\subsection{Technologies for Reducing Stigma}
\label{sec: external factors}
\begin{figure}
    \centering
    \includegraphics[width=0.8\textwidth]{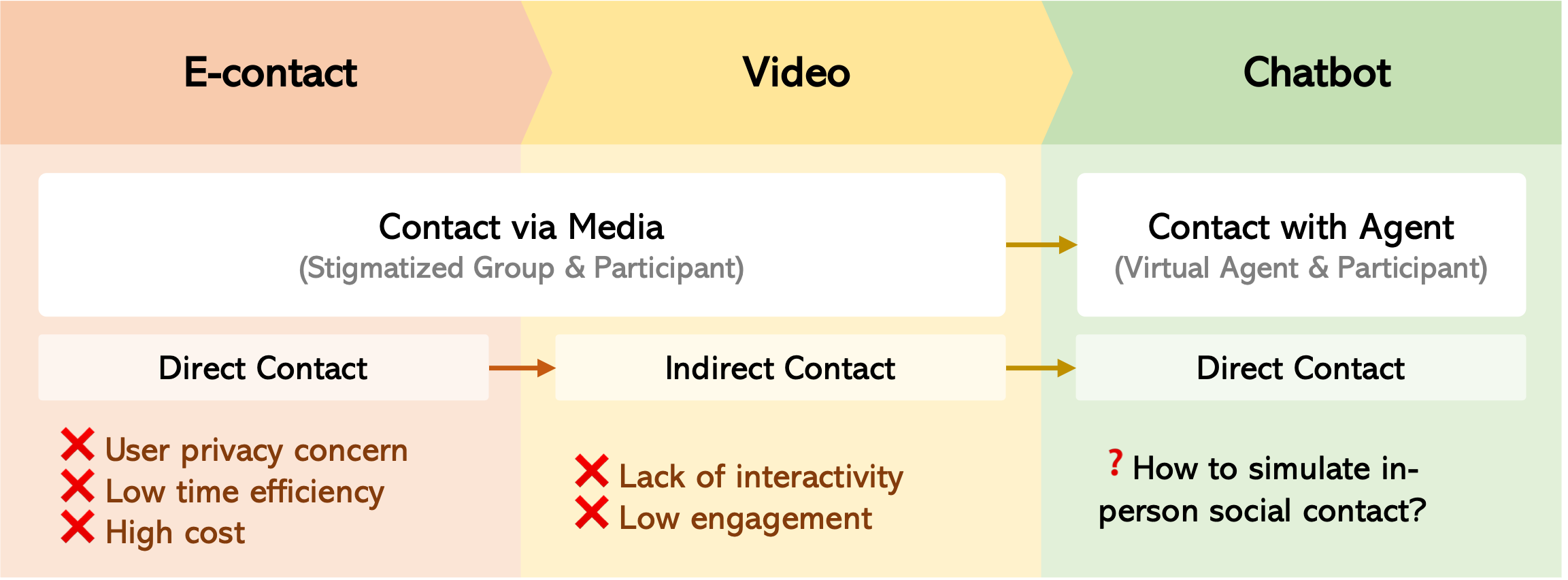}
    \caption{
    Recent studies leverage digital technologies for social contact interventions. In one scenario, technology serves as a medium facilitating human-to-human contact. In another case, technology functions as a virtual agent representing the stigmatized group, engaging directly with participants.
    }
    \label{fig:tech-for-stigma}
\end{figure}

Previous studies have demonstrated that technology-facilitated social interactions—whether in-person, remote, or simulated—can effectively diminish the stigma associated with mental illness \cite{corrigan2007will, sartorius2010wpa, li2013evaluation, brown2020effectiveness, pendse2021can, stelzmann2021can, rodriguez2022innovative}, as shown in Figure \ref{fig:tech-for-stigma}. Such interactions through technology encourage individuals with mental health conditions to openly share their experiences, foster their close and supportive relationships with others, and enhance public understanding of mental health issues \cite{corrigan2002understanding, evans2012mass}. 
For instance, engaging with mental health service users through virtual reality and communication technologies has been successful in reducing mental-health stigma among members of the public \cite{rodriguez2022innovative}.

Despite these efforts, technology-mediated solutions face significant challenges in scaling up usage. For example, e-contact interventions \cite{pendse2021can}, in which members of different groups never meet physically but interact online, risk exposing individuals to stigmatization and demand substantial time and effort from participants \cite{sartorius2010wpa}. By contrast, approaches where people watch recorded regular~\cite{brown2020effectiveness} or VR~\cite{stelzmann2021can} videos are cost-effective, but are generally only unidirectional exposure to out-group member(s), and lack real-time interaction and engagement.

To tackle the above challenges, researchers have proposed employing virtual agents that simulate the stigmatized group for social contact with individuals.
Due to their cost-effectiveness and constant availability, conversational agents—colloquially known as chatbots—have considerable potential to mitigate public stigma by simulating social interactions \cite{grove2021co, skjuve2018chatbots}, improving users' understanding of mental health concerns \cite{grove2021co, skjuve2018chatbots}, facilitating humans’ in-depth self-disclosure of thoughts and feelings \cite{lee2020hear}, increasing their engagement in communication \cite{skjuve2022longitudinal, de2020effectiveness}, and driving behavioral change by helping people reflect on and learn from their own experiences \cite{kocielnik2018designing, wang2020alexa, lee2019caring, bickmore2013automated}.

Expanding on these findings, a recent investigation concluded that a narrative chatbot could potentially reduce stigmatizing thoughts \cite{lee2023exploring}. More specifically, the study found a relationship between interacting with such a chatbot and two factors: 1) a shift in users' existing belief that individuals are personally responsible for their mental health issues and 2) an increase in users' inclination to offer assistance. Evaluation of the outcome involved prompting users to assign responsibility to internal factors (such as personal reasons) or external factors (such as genetic or environmental factors) and determining if individuals expressed the intention to make personal changes \cite{corrigan2003attribution, corrigan2013erasing}.
While this study is a starting point for exploring chatbots' effects on reducing stigma, it focused
on chatbots delivering narratives about depression.
It did not consider more diverse interaction dynamics that are possible between humans and chatbots, and did not evaluate user experience, which is a critical aspect of designing effective chatbots.   
Our study builds on that prior work by examining the effect of more complex interaction designs (i.e., human-chatbot cooperation) on stigma reduction and evaluating the impact of these designs on user experience.

\subsection{Human-Chatbot Cooperation for Changing Attitudes}

The concept of human-artificial intelligence cooperation (HAIC), is attracting increasing research interest as AI’s emotional, behavioral, and cognitive abilities continue to advance \cite{arous2020opencrowd, cai2019hello, oh2018lead, wang2019human}. 
By functioning as an autonomous and unique contributor to a team, HAIC can usefully inform the construction of chatbots aimed at capturing the benefits of cooperation as a means of combating stigma.

HAIC studies have begun exploring the impact of different HAIC modalities on people’s perceptions and attitudes \cite{ashktorab2020human, bezrukova2023artificial, li2022human, yue2023impact, hou2023exploring}. For instance, Ashktorab et al. \cite{ashktorab2020human} investigated how their participants’ beliefs about their cooperative gaming partners being either human or AI influenced their social perceptions and game performance. In this study, impressions of chatbots (e.g., intelligence) were utilized to create a measurement instrument for assessing factors contributing to favorable collaboration outcomes \cite{tijunaitis2019virtuality, oliver2019communication}.
Yue et al. \cite{yue2023impact} conducted a study in the field of customer service to investigate how different types of human-AI cooperation and expected outcomes can affect consumers' willingness to use. While previous research has identified potential factors (e.g., AI's role \cite{yue2023impact}, confidence \cite{ashktorab2020human}, personal characteristics \cite{hou2023exploring}, etc.) that can influence people's attitudes during human-AI cooperation, there is limited knowledge on whether cooperation or collaboration itself can bring about changes in human attitudes. Furthermore, although there are studies that examine the use of storytelling to influence stigmatized attitudes \cite{lee2023exploring}, there is a noticeable research gap in understanding the impact of cooperation on people's stigmatized attitudes, which the present study aims to fill.

\subsection{Research Questions}

We propose three different chatbots to simulate the three interaction designs, considering their interaction modes and content topics. We expect that diverse interactions between AI and humans will yield varying effects on two aspects: 1) participants' impressions of chatbots and perceptions of mental health, and 2) changes in stigmatizing attitudes towards it. 
By understanding how interaction modes and content topics impact both user impressions and stigma, we can develop interventions that are not only effective but also engaging and positively received by users. 
Moreover, Allport has noted in Contact Hypothesis \cite{allport1954nature} that only when users' experiences with outgroup members are “positive” can social contact effectively reduce anxiety between conflicting groups and thereby change their attitudes toward stigmatized groups, which suggests a correlation between the two aspects. 
Accordingly, our research questions (RQs) are as follows.

\begin{itemize}
    \item \textbf{RQ1.} 
    What is the impact of different interaction mode and content topics on changing users' impressions about the chatbot?
\end{itemize}

We address this question by examining participants' survey responses, interviews, and message content during their conversation with the chatbot. 
Furthermore, we are also interested in the effects of interaction process on influencing users'
stigmatized thoughts, which brings us to our next RQ:

\begin{itemize}
    \item \textbf{RQ2.} 
    How, if at all, do different interaction modes and content topics affect users’ stigmatizing thoughts about mental illness?
\end{itemize}

We address RQ2 by conducting a comparative analysis of pre- and post-surveys. Additionally, we examine the interview data and compare our results with relevant theories and previous empirical research outcomes \cite{dobson2022myths}. This methodology allows us to examine the cooperative relationship in accordance with Allport's  \cite{allport1954nature} hypothesis that equal status, a common goal, and inter-group cooperation are key factors for reducing prejudice, in order to understand the underlying factors that contribute to the observed changes. 

By addressing both RQ1 and RQ2, we aim to provide a comprehensive understanding of the potential of HAIC in reducing mental illness stigma in a sensitive context. 
\section{Methods}

Based on our prior chatbot-design experience and review of the relevant literature, we expect that a chatbot we create can meet the requirements of the Intergroup Contact Hypothesis and thus reduce its human interlocutors’ stigmatizing thoughts about mental illness. Specifically, our finalized chatbot assumed the persona of a university student, \textit{Holly}, who had first-hand experience of depression.

To unravel the roles of cooperation tasks and mental illness-related content, we designed three variations of the chatbot. We evaluated them through a mixed-methods experiment in which participants participated in a learning activity with one of the chatbots, as summarized in Table~\ref{tab:participant_groups}. By designing three groups, we simulated the three different designs in social-contact interventions in order to address RQ1 and RQ2.

Our study design excluded a  'non-cooperative × other topics' condition, since both theoretical and empirical evidence suggest it would not effectively reduce stigma. 
According to the Contact Hypothesis \cite{allport1954nature}, effective stigma reduction relies on positive, cooperative interactions that foster mutual understanding and goal alignment, 
which would not be served by such a condition.  
Moreover, prior reseach \cite{desforges1991effects, dansereau1988cooperative, kohrt2021collaboration} has concluded that such a condition is not effective at reducing stigma in human-human social contact, thus there is no evidence that it should be effective in human-agent interactions either. With no evidence supporting its effectiveness in human-agent contexts, we excluded this condition from our study, as it would likely be ineffective across both interaction types.
Our study was approved by our University's Institutional Review Board (IRB).

\small 
\begin{table}
\caption{Participants were organized into three groups, which varied in whether they engaged in a one-way information dissemination task or a two-way cooperation task with the chatbot, and whether the learning materials were related to mental illness. The 'References' column lists prior research corresponding to each group's setup.}
\label{tab:participant_groups}
\centering
\begin{tabular}{lllll} \toprule
\textbf{Group name} & \textbf{User's task} &  \textbf{Interaction mode} & \textbf{Content topics} &  \textbf{References}         \\ \midrule
Group 1  & Reading &  Information dissemination    & Related to mental illness & \cite{lee2023exploring, rodriguez2021controlled, stelzmann2021can, yuen2021effects, brown2020effectiveness}   \\
Group 2   & Cooperation & Cooperation task   & Related to mental illness & \cite{ho2017reducing, kohrt2021collaboration}    \\
Group 3   & Cooperation & Cooperation task   & Unrelated to mental illness & \cite{desforges1991effects, maunder2019modern} \\ \bottomrule
\end{tabular}    
\end{table}

\subsection{Participants}
Study participants were recruited from local universities via social media networks and online announcements. The selection criteria required participants to be "students over 18 years old". Although university students do not fully represent the general population, they offer a diverse and accessible demographic that is generally familiar with technology and can provide meaningful insights into the use of chatbots for addressing mental health stigma.
The participants were informed that there would be one task per day, each taking between 10 and 20 minutes; and that those who completed all assigned tasks over the two weeks of the experiment and submitted the required surveys would receive compensation of US\$85. 
On the other hand, those who opted to discontinue their participation during the study would receive compensation at the rate of US\$6 per task completed. Only those who completed all tasks were eligible to attend the interview, for which they would be granted an additional payment of US\$12. All participants were also assured that if any study question or other content caused them discomfort, they were free to skip it without incurring any penalty.

The recruitment process yielded 84 participants, of whom 50 identified as female and 33 as male, with one unwilling to disclose their gender. Their average age was 23.2 (SD=2.7). None of the participants reported a current mental illness or attendance at counseling sessions. Each was assigned a unique identification number, and the range of these numbers was reflective of their group membership. Specifically, Group 1 included participants with identification numbers P1-P26; Group 2, those with numbers P27-P55; and Group 3, those with numbers P56-P84.
To ensure balance among the three groups and to control for the potential impacts of mental-health literacy and gender on mental illness stigma, participants’ scores on the SDS and gender distribution were considered. Accordingly, Group 1 included 17 female participants, Group 2, 16, and Group 3, 18. The whole sample’s mean SDS score was 11.40 (SD=3.83).

\subsection{Procedure}
All participants signed their consent forms in the beginning. They were presented with a pre-survey and vignette aimed at capturing their initial attitudes toward a person with a mental illness. After this, the participants underwent a pre-training task to familiarize them with interacting with our chatbot. In it, the chatbot introduced itself as \textit{Holly} but explicitly conveyed its non-human nature. Further, it clarified to the participants that their responses would be shared exclusively with the researchers and not disclosed to other participants. Only after the pre-training session were participants eligible to proceed to the main study.

Throughout the two-week experiment, participants received daily task reminders from the chatbot. Each group was allotted 16 hours (i.e., from 8 a.m. until 11:59 p.m.) to complete each day’s task, which was designed to last 15-20 minutes. If a participant attempted to access the chatbot outside of the designated daily task window, the chatbot would not respond.
Given the challenge of keeping users engaged in a longitudinal study, we allowed them some flexibility about missing tasks. However, if a participant missed tasks on more than two consecutive days, their responses were deemed invalid, and they were advised to discontinue their participation, with partial reimbursement provided as explained above. 
In the end, 6 out of 84 participants failed to meet our criteria by missing more than 6 out of the 14 tasks. 
This resulted in a valid participant sample size of 78, with 25 participants in Group 1, 27 in Group 2, and 26 in Group 3. 
These data were utilized for our further data analysis.

\begin{figure}
    \centering
    \includegraphics[width=\textwidth]{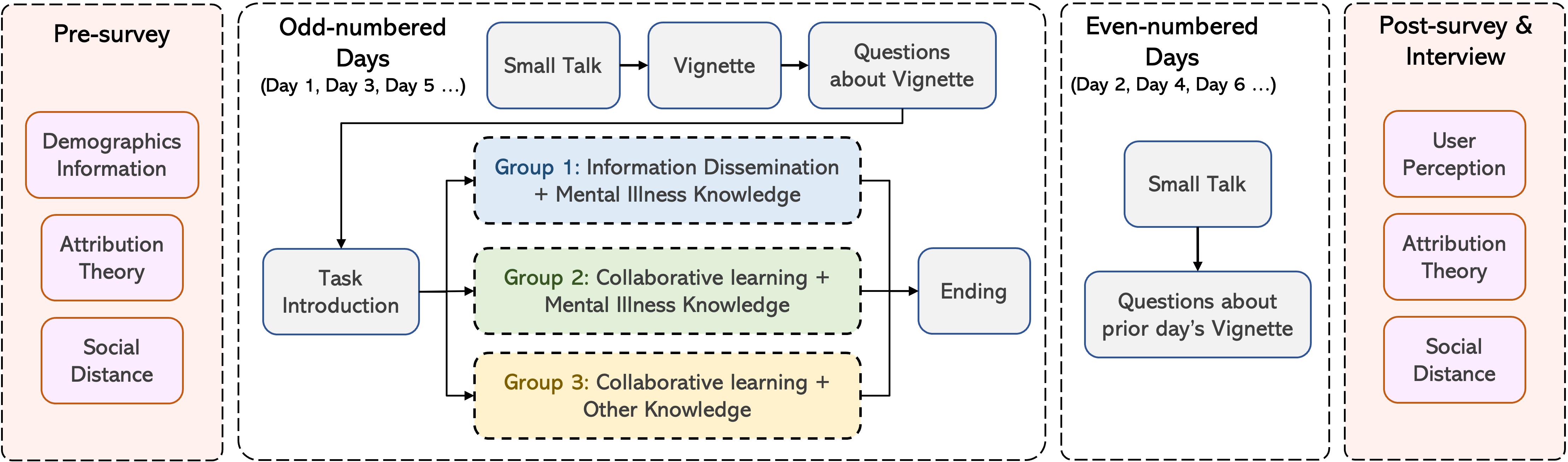}
    \caption{
    Experimental procedure.
    Throughout the two-week study period, all three groups completed daily tasks using a Telegram chatbot, \textit{Holly}. 
    On odd-numbered days, participants engaged in small talk, were presented with a vignette about \textit{Holly}'s experiences related to mental illness, and then engaged in a learning task.
    For Group 1, this task consisted solely of reading a message aimed at boosting their mental illness knowledge. Groups 2 and 3, on the other hand, engaged in a cooperation task to absorb the study content. 
    The difference between Groups 2 and 3 is that Group 2's learning content was related to mental illness, while Group 3's learning content was not.
    On even-numbered days, participants in all three groups engaged in small talk and answered questions posed by the chatbot.}
    \label{fig:study-design}
\end{figure}

Our study design was derived based on previous literature on long-term human-chatbot interaction \cite{hobert2020small, lee2020hear, skjuve2022longitudinal, lee2023exploring} and face-to-face social contact task design \cite{desforges1991effects, rodriguez2021controlled, lee2023exploring}. Specifically, the design of the odd-even day structures and vignettes was derived from the previous long-term human-chatbot interaction \cite{lee2020hear, skjuve2022longitudinal, lee2023exploring}, and the interaction task (i.e., cooperation or information dissemination) was derived from previous social contact designs \cite{desforges1991effects, rodriguez2021controlled, lee2023exploring}.

As shown in Figure \ref{fig:study-design}, on odd-numbered days (e.g., Days 1, 3, 5, 7, etc.), daily interaction commenced with a warm-up session that included 
small talk, the chatbot sharing a first-person vignette~\cite{bickmore1999small}, and then asking questions about the vignette.
Some of these questions were intended to prompt users’ recollections of their own life events, e.g., \textit{“Have you ever had similar experiences?”} and \textit{“Do you have any suggestions for my situation?”} 
The vignettes were adopted from previous literature that demonstrated their effectiveness in increasing user engagement \cite{lee2023exploring}. Specifically, they consisted of short first-person stories about \textit{Holly}’s experiences with \textbf{mental illness} across seven scenarios: 1) academic study, 2) working, 3) intimate relationships, 4) interactions with friends, 5) interactions with family, 6) interactions with strangers, and 7) being alone.
For instance, the vignette about work was as follows: \textit{"When I’m at work, I get things done as usual—but I know that it is not my best. In the past, I could take care of 5-6 tables of customers at the same time; however, now I can only manage 1 table...”}. 
The chatbot shared these vignettes across short messages of no more than two sentences and did not continue until the user made some kind of response. The vignettes’ design was aimed at ensuring that the participants gained a comprehensive understanding of how \textit{Holly} was affected by mental illness. 
Following the warm-up session, participants were expected to carry out their individual learning tasks with the chatbots. Lastly, during the "Ending" of each odd-day, participants were asked about any insights they had gained from the day’s materials.

On even-numbered days, \textit{Holly} presented all participants with additional questions about the previous day's vignette, aiming to capture the potentially different effects of our three task approaches on participants’ perceptions and attitudes. Specific follow-up questions included \textit{“Imagine if you are my classmate, I wonder if you would want to be my teammate for school projects?”}, \textit{“Do you think I could still be as successful as other colleagues?”}, and \textit{“Do you agree that finding love when you have an illness is impossible?”}

Upon the experiment’s conclusion, the participants were asked to complete a post-survey containing the same items as the pre-survey. This enabled us to examine the extent to which the two-week interventions had influenced their perceptions of and attitudes toward individuals with mental illnesses. Unlike the pre-survey, however, the post-survey did not include a vignette or any similar narrative. Rather, responses were to be based on the perceptions of \textit{Holly} that the subjects had accumulated over the two weeks of their participation.

While recruiting interviewees, efforts were made to maintain balanced numbers from each group. Among the 41 interviewees, representing 53\% of all study participants, there were 14 from Group 1; 14 from Group 2; and 13 from Group 3. At the conclusion of the study, the researchers debriefed each participant on the research objectives and addressed any concerns that were raised. Ethical approval for this research was granted by our university’s Institutional Review Board.

\subsection{Task Design}

\begin{figure}
    \centering
    \includegraphics[width=0.8\linewidth]{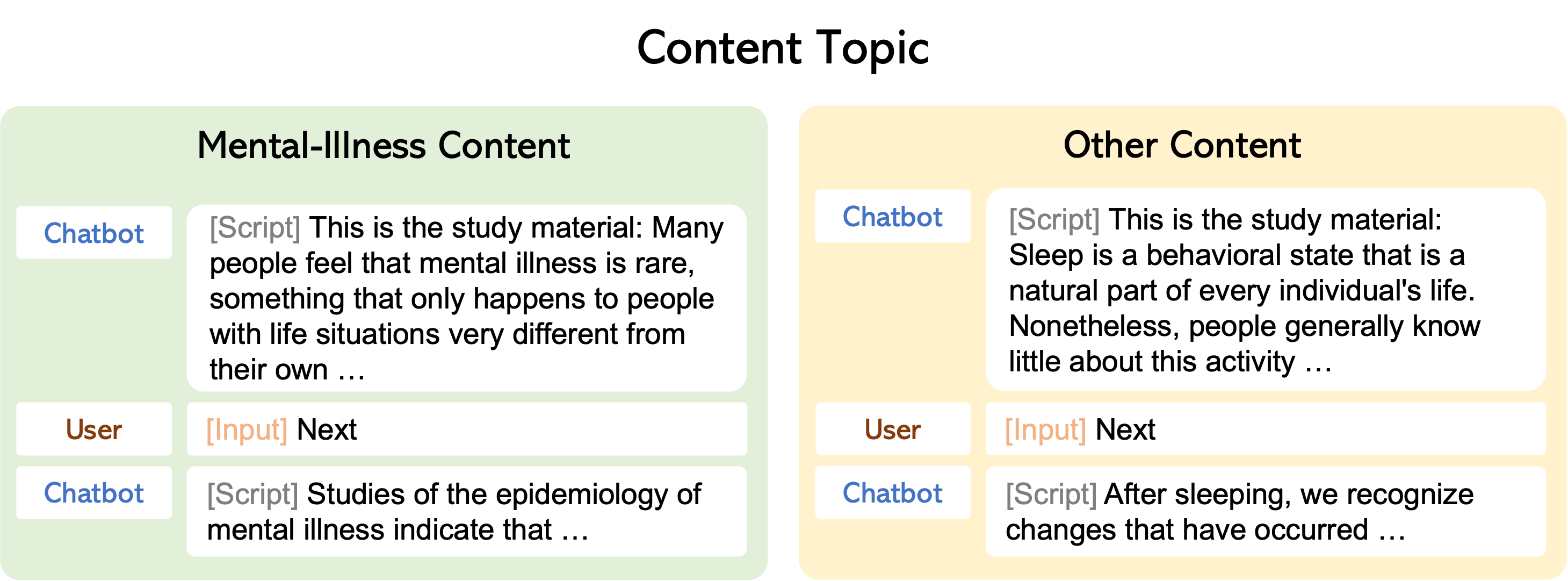}
    \caption{Illustrative example of content topic designs.}
    \label{fig:content-topic}
\end{figure}

In our study design, all three groups commenced by participating in identical small talk and vignette sessions, then undertook a learning task that was unique to each group. Two types of content topics were chosen for the learning task, as shown in Figure \ref{fig:content-topic}. Groups 1 and 2, which received mental illness content, were presented with identical learning materials taken from {\it "Information about Mental Illness and the Brain"} by the U.S. National Institutes of Health (NIH) \footnote{https://www.ncbi.nlm.nih.gov/books/NBK20369/}, mirroring the educational content used in a previous study aimed at reducing stigmatizing thoughts \cite{watson2004changing}. Group 3, on the other hand, received study content unrelated to mental illness: i.e., {\it "Information about Sleep"}, also from the NIH. Both types of content came from the same institutes in order to control their style of expression and content quality. In addition, we controlled the length of the content, keeping it within a range of 110-140 words per day.

\begin{figure}
    \centering
    \includegraphics[width=0.8\linewidth]{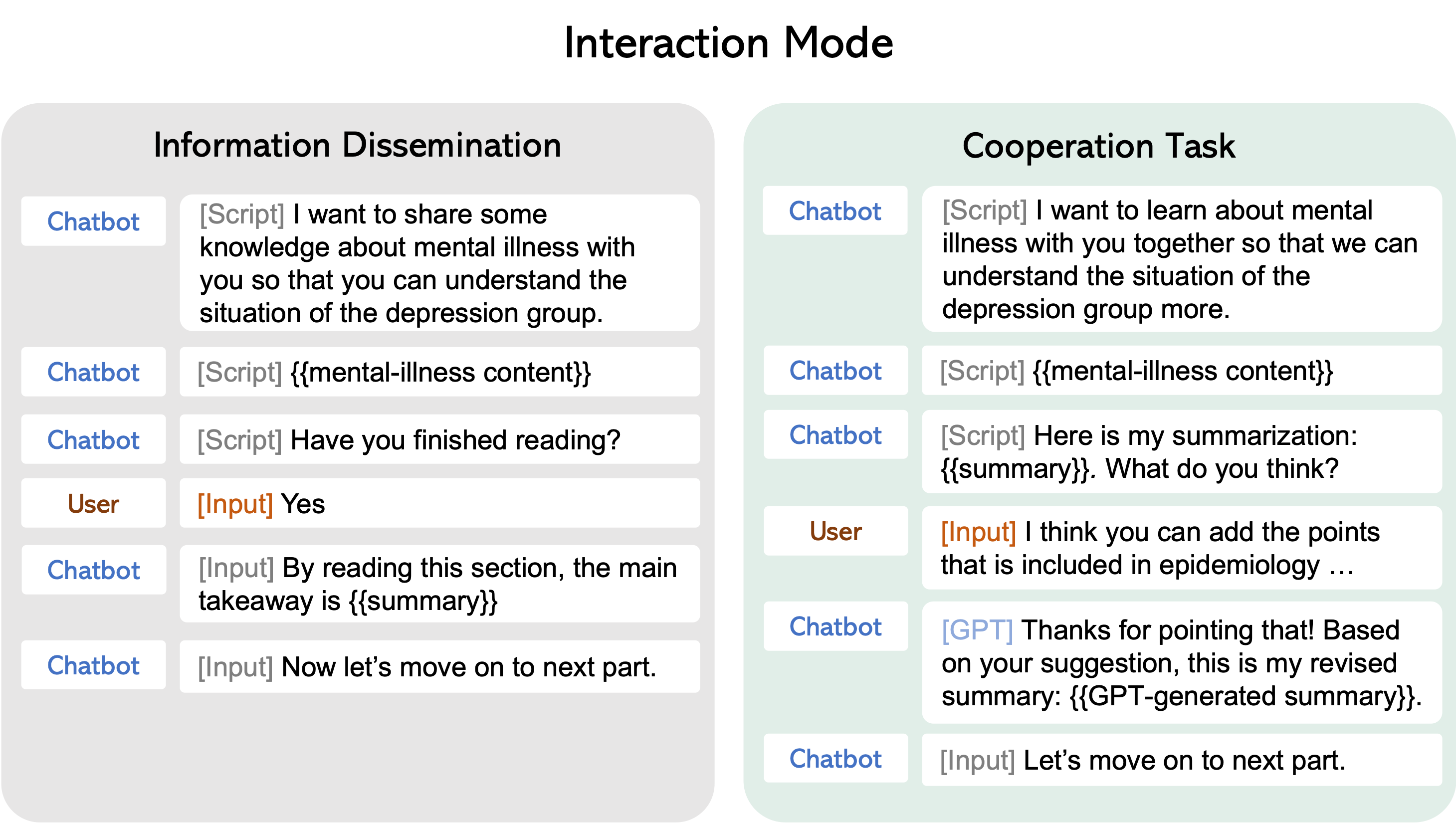}
    \caption{Illustrative example of interaction mode designs.}
    \label{fig:interaction-mode}
\end{figure}

The difference between the two interaction modes ("information dissemination" and "cooperation task") is shown in Figure \ref{fig:interaction-mode}. Group~1 simulated the information dissemination mode, where the chatbot directly shared the content and the takeaway, and Group~1 participants didn't need to take any further action.
In contrast, participants in Group~2 and Group~3 performed a cooperation task, where they and the chatbot collaboratively created and corrected summaries of the learning content. The tone of the chatbot was controlled to be friendly in both conditions. Average daily interaction time was slightly longer in Groups 2 and 3 than in Group 1, due to the users' typing time during the cooperation task (Group 1: Mean=10.16 minutes, STD=3.77; Group 2: Mean=13.64 minutes, STD=5.36; Group 3: Mean=13.18 minutes, STD=4.05).

\begin{figure}
    \centering
    \includegraphics[width=0.75\textwidth]{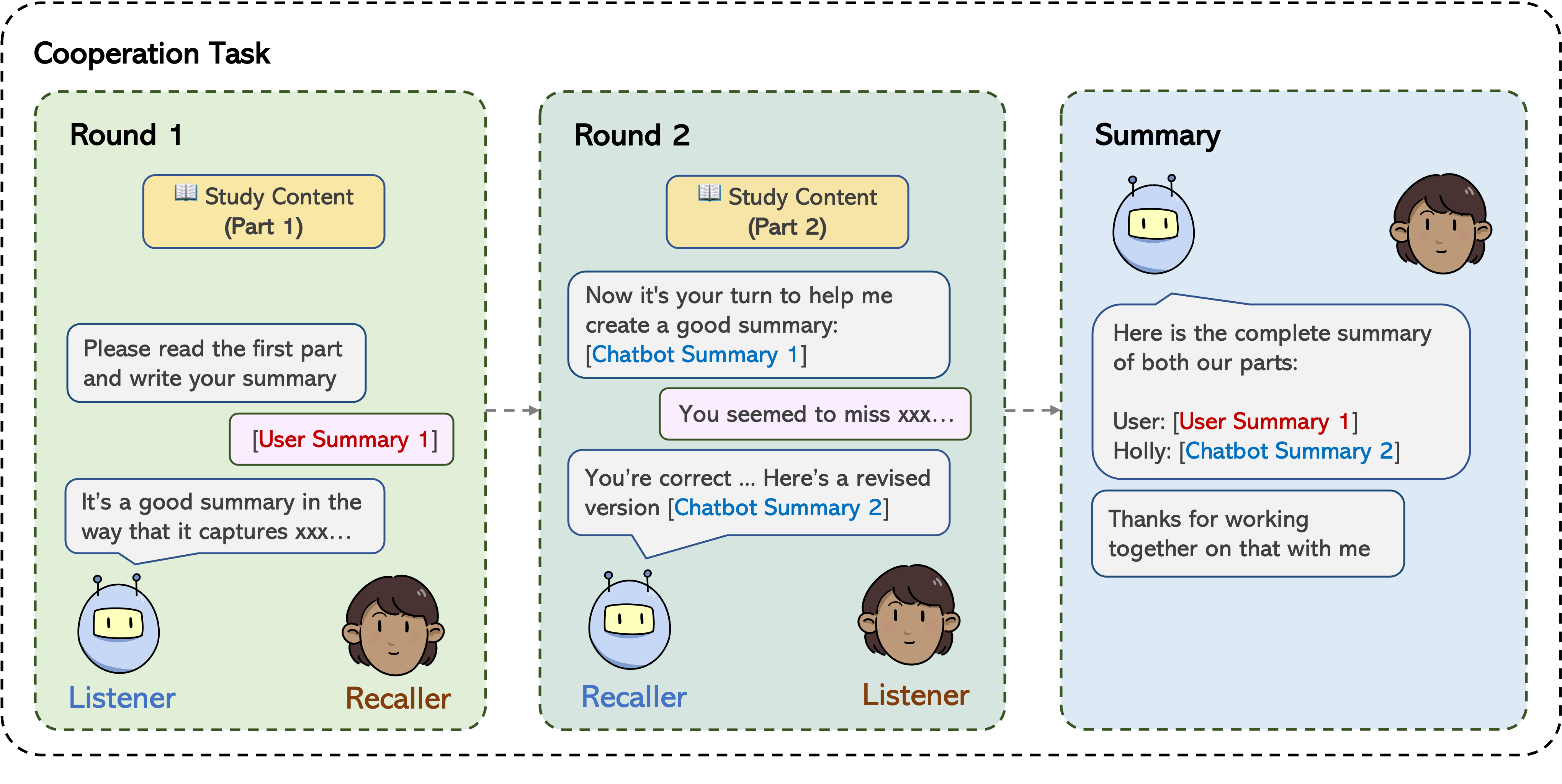}
    \caption{Each cooperation task was comprised of two rounds. Within each, the chatbot and the human user alternated between producing a summary of the learning material and correcting the summary provided by the other party. Following the completion of two rounds, the chatbot presented the user with the final summary.}
    \label{fig:task-design}
\end{figure}

Drawing inspiration from previous studies showing the promise of cooperation interventions for addressing social stigma~\cite{desforges1991effects, walker1998academic, dansereau1988cooperative},
we have defined two distinct roles within the cooperation process: \textbf{listener} and \textbf{recaller}. Both roles access an identical segment of study materials and have enough time to read and comprehend them. The recaller tries to summarize the article's essence while the listener offers corrections for any inaccuracies. After one round, the recaller and listener switch roles and continue. In our design, as depicted in Figure \ref{fig:task-design}, on Days 1, 5, 9, and 13, we allocated the first-round roles of listener and recaller to the chatbot and the human participant, respectively. On Days 3, 7, and 11, on the other hand, the chatbot started as the recaller and the human as the listener in the first round. Our intention was to mitigate any potential impact of sequence on the subjects’ interpretations of the study process. Throughout each round of summarization and corrective feedback, the chatbot provided clear instructions regarding its designated role and associated responsibilities, as detailed below.

\begin{itemize}
    \item \textbf{Listener:} In the role of listener, the chatbot prompts its user to create summaries. Following the user’s submission, the chatbot uses the GPT-3.5 API to compare the summary against the original NIH paragraph. Once that process establishes that the initial summary was adequate or inadequate, the chatbot will tell the user so, and in cases where it was inadequate, also prompt the user to rephrase the summary, e.g., by saying \textit{“Please incorporate my feedback into your summary and include your previous content in its entirety”}. Once it receives a revised summary, the chatbot commends it without requesting further modifications, regardless of its quality.

    \item \textbf{Recaller:} When operating as the recaller, the chatbot first generates a summary of the NIH materials. It then invites the user to provide feedback on the generated summary. In cases where the user offers suggestions for improvement, the chatbot revises its response and regenerates it accordingly.
\end{itemize}

\subsection{System Implementation}

\begin{figure}
    \centering
    \includegraphics[width=\textwidth]{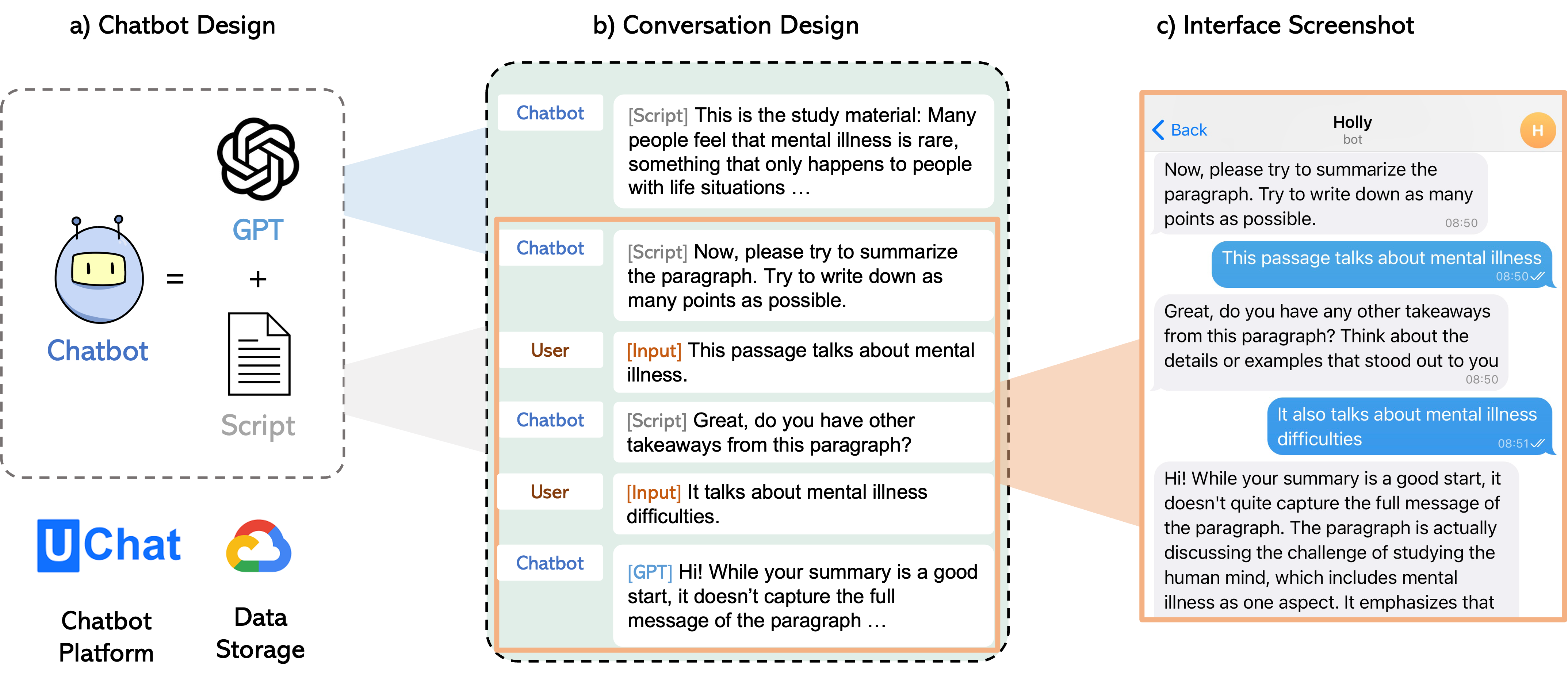}
    \caption{System design of the chatbot. (a) shows the chatbot development architecture, where the chatbot was deployed on Uchat, supporting script-based and GPT-based responses. (b) was an example of conversation flow, including script-based and GPT-based conversations. (c) was a screenshot of the Telegram interface that users interacted with.}
    \label{fig:system-design}
\end{figure}

Figure \ref{fig:system-design} illustrates the implementation of our chatbot, \textit{Holly}. We developed \textit{Holly} using Uchat\footnote{https://uchat.au/}, a platform that facilitates chatbot creation through both script-based and AI-integrated responses. Script-based messages were employed to establish the main conversational flow, while AI-integrated responses (GPT-3.5) were used to enhance the naturalness of the interactions. GPT-3.5 was chosen since it was a state-of-the-art language model at the time of the study.

To ensure consistency in Holly's tone, we designed a uniform system message - \textit{"You are a college student named Holly, who has undergone a tough situation in depression. Give friendly feedback to the user. Talk in a friendly and concise style. Give a response of less than 40 words."} - for each GPT-3.5 API call and tailored specific context prompts accordingly.  
We tested these prompts during a pilot study to ensure their effectiveness. Additionally, we collected daily feedback from users to monitor the appropriateness of the generated content. The user feedback indicated that the overall quality of the AI-generated responses was appropriate and satisfactory. 
To further ensure content quality, we conducted a manual review of the conversation logs between participants and chatbots. The review showed that the responses were generally well-suited to the topic, natural in context, and free from stigmatizing content. However, occasional system errors were noted, including duplicate messages (e.g., sending the same message twice, observed 13 times [1.1\% of conversations]) and lapses in persona (e.g., responses like "As an AI, I do not have attitudes towards...", observed three times [less than 0.2\% of conversations]).
Although these issues were anticipated and addressed in the prompt design, they still occurred infrequently. We report them here for transparency and clarity of the study's findings.
The precise system and context prompts are detailed in the Appendix \ref{app:prompts}.

Using Uchat, we deployed \textit{Holly} on Telegram, enabling participants to interact with the chatbot from their mobile devices. Additionally, Uchat supports external data storage, and we used Google Cloud to store the participants' conversation data. Before the study began, all participants were informed that their interactions would be recorded and shared with the research team.

\subsection{Measurements}

To answer our RQs, we triangulated our analysis across the survey responses, daily conversational logs, and interview data, as detailed below.

\subsubsection{Survey}

The survey items were adapted from three questionnaires in the prior literature: User Impression \cite{ashktorab2020human}, Corrigan et al.’s \cite{corrigan2003attribution} Attribution Questionnaire, and the SDS \cite{link1987social, norman2008role}. The full content of survey items can be found in Appendix \ref{app:survey}.

{\bf User Impression.} 
User impressions of \textit{Holly} were measured using an instrument adapted from previous human-AI interaction research, where it was used to understand how social attitudes impact confidence in AI abilities and cooperative performance \cite{ashktorab2020human}. 
The construct of user impressions encompasses four key dimensions: intelligence, rapport, likability, and creativity. 
Each of these dimensions was gauged through the following items, using 7-point ordinal scales.

\begin{itemize}
\item \textbf{Intelligence:} Intelligence was assessed through four items: Unintelligent/Intelligent, Ignorant/Knowledgeable, Incompetent/Competent, and Irresponsible/Responsible. 
The cumulative intelligence score is derived from the mean of these four scales.
\item \textbf{Rapport:} Rapport was measured by asking participants to rate statements such as "Holly appeared engaged" or "Holly and I collaborated towards a shared objective."  A rapport score was calculated by taking the mean of nine such items on a scale from ``strongly disagree'' to ``strongly agree.'' 
\item \textbf{Likeability:} To measure likeability, participants rated the chatbot along the dimensions of unfriendly/friendly, not kind/kind, unpleasant/pleasant, not cheerful/cheerful, and dissimilar to me/similar to me.
\item \textbf{Creativity:} Lastly, the assessment of creativity involves the use of three semantic differential scales: funny/funny, not creative/creative, and unique/ordinary.
\end{itemize}

{\bf Social Distance.} The SDS is widely employed to assess the respondents’ inclinations towards specific behaviors involving others grappling with mental health disorders \cite{baumann2007stigmatization, corrigan2003attribution}. It comprises seven items that probe various levels of personal and social proximity to an individual like \textit{Holly}, who represents a person with a mental health disorder. Examples of these items include: \textit{“How would you feel about renting a room in your home to someone like Holly?”} and \textit{“How would you feel about recommending someone like Holly for a job working for a friend of yours?”}. Other items explore scenarios such as working alongside \textit{Holly}, having \textit{Holly} as a neighbor, trusting \textit{Holly} to take care of one's children, the prospect of \textit{Holly} marrying into the family, and introducing \textit{Holly} to friends.
Responses to all SDS items are given on a four-point scale ranging from 0=\textit{“definitely willing”} to 3=\textit{“definitely unwilling”}, which means that the total SDS range is from 0 to 21, with lower numbers indicating less stigma.

{\bf Attribution.} The attribution-related items within the questionnaires \cite{corrigan2000mental} encompassed three primary dimensions: 1) Beliefs Pertaining to Personal Responsibility, 2) Emotional Responses, and 3) Behavioral Reactions. The first consisted of a single subdimension, Blame, whose items were intended to capture the respondents’ beliefs about the extent to which \textit{Holly} was personally responsible for her mental illness.

The Emotional Responses dimension had four subdimensions. \textit{Pity} measured the extent to which the participants experienced sympathy towards \textit{Holly}. \textit{Anger} measured the level of anger participants felt towards \textit{Holly}. \textit{Fear}, assessed their perception of danger and threat associated with \textit{Holly}. Lastly, \textit{Dangerousness} measured the extent to which participants believed that \textit{Holly} represented a safety risk.

The Behavioral Reactions dimension also comprised four subdimensions. \textit{Help} evaluated participants’ willingness to support people with mental illness like \textit{Holly}. \textit{Avoidance} measured the extent to which they would prefer to stay away from \textit{Holly}. \textit{Coercion} measured their willingness for medication management and/or other treatments to be forced upon \textit{Holly}. \textit{Segregation} measured the participants’ willingness to keep \textit{Holly} away from her community.

All Attribution items were answered on a nine-point semantic differential scale from 1=\textit{“not at all”} to 9=\textit{“very much”}.

\subsubsection{Interviews}

Our interviews followed a semi-structured format and lasted 35 to 45 minutes. Their primary topics of focus were the interviewees’ 1) engagement in daily activities involving the chatbot, 2) perceptions of the chatbot’s operational efficacy within the cooperation tasks, 3) impressions of and attitudes toward the \textit{Holly} persona as depicted in the vignettes, and 4) reflections on whether and how their study involvement affected their perceptions of individuals with mental health issues. Questions on the first and second interview topics included asking about the participants’ familiarity with chatbot usage, their study habits, and their understanding of mental health illness. Questions covering the third and fourth topics prompted participants to describe their perspectives on \textit{Holly} (including her task performance, in the case of interviewees drawn from Groups 2 and 3), as well as their reasons for forming those impressions. Specifically, participants were asked about their perceptions of the chatbot's knowledge and competence, which are common variables measured in studies relevant to human-agent cooperation \cite{yiren2023warmth, mckee2022aamas, ashktorab2020human}. 
Additionally, the interviewees were asked about any changes in these impressions over time and the reasons underlying such changes.

We audio-recorded the interviews with the participants’ permission, transcribed them, and used thematic analysis to sort the responses into relevant categories based on the questions’ contexts \cite{nowell2017thematic}. At the start, two researchers independently reviewed all the interview data multiple times and classified eight participants’ answers to create preliminary coding structures. Afterward, the same two researchers separately coded the remaining interview answers, then met to compare and reconcile their coding results, with changes being added iteratively until both agreed on the coding system and inter-rater reliability had reached an acceptable level ($\kappa = .85$).

\subsubsection{Conversational Logs}
The participants underwent a two-phase questioning process over a two-day period about each vignette they had read. Initially, they were asked if their own experiences aligned with the situation described in the vignette. If they answered in the affirmative, they were then asked to provide details. In the next phase, they were prompted to share their thoughts and feelings about the actions of the \textit{Holly} character and asked to give advice about her situation described in the vignette from the previous day.

To examine inter-group variations in responses, two raters independently analyzed all the collected data ($\kappa = .87$). Before conducting that formal evaluation, they practiced by rating the Day 1 and Day 3 responses, then deliberated on any disparities until they achieved consensus and refined their approach accordingly. 

In this process, responses were categorized based on whether the participants exhibited empathetic reactions during the conversation \cite{sharma2020computational}. For instance, a statement like {\it “I’m sorry to hear this”} effectively conveyed empathy and qualified as an empathetic response. 
In each of the three experimental groups, the total number of responses within a span of 14 days was counted, and then divided by the number of responses that contained empathetic reactions to arrive at the group ratio of emotional reactions. This was done because of prior literature’s suggestion that empathetic reactions play an important role in establishing empathetic rapport and support \cite{elliott2011empathy}. Thus, empathetic expressions within our participants’ discourse could indicate positive social interaction with the chatbot, thereby contributing to the answer to RQ1 and RQ2.

\section{Results}

\subsection{Effects of Interaction Design on Users' Impressions of the Chatbot (RQ1)}

We answered RQ1 by examining the survey results, conversational log, and interview data from three groups. In particular, we conducted a comparative analysis of the experiences within these groups to understand how different interaction designs affected their impressions of chatbots, as explained below.

\subsubsection{Overall impressions of the Chatbot}

We performed a normality test (Shapiro-Wilk), which showed that our data did not fit the normal distribution, so we used the Kruskal-Wallis test, which is a nonparametric method for single-factor analysis, to analyze the survey data, focusing on user impressions measured by the dependent variables (DVs), with group membership as the independent variable (IV).
Our findings revealed significant group membership effects in \textbf{intelligence} (F=9.67, p<.01) and \textbf{likeability} (F=13.13, p<.01). However, no significance was observed for rapport and creativity.

We conducted a post-hoc analysis using Dunn's Test, which revealed significant differences in intelligence between Group 1 (M=4.98, SD=0.95) and Group 2 (M=5.59, SD=0.83; p<.05), as well as between Group 1 and Group 3 (M=5.51, SD=0.70; p<.05, Bonferroni adjusted). Similarly, in terms of likeability, significant differences were found between Group 1 (M=4.61, SD=0.82) and Group 3 (M=5.28, SD=0.68; p<.01, Bonferroni adjusted).

\begin{figure}
    \centering
    \includegraphics[width=\textwidth]{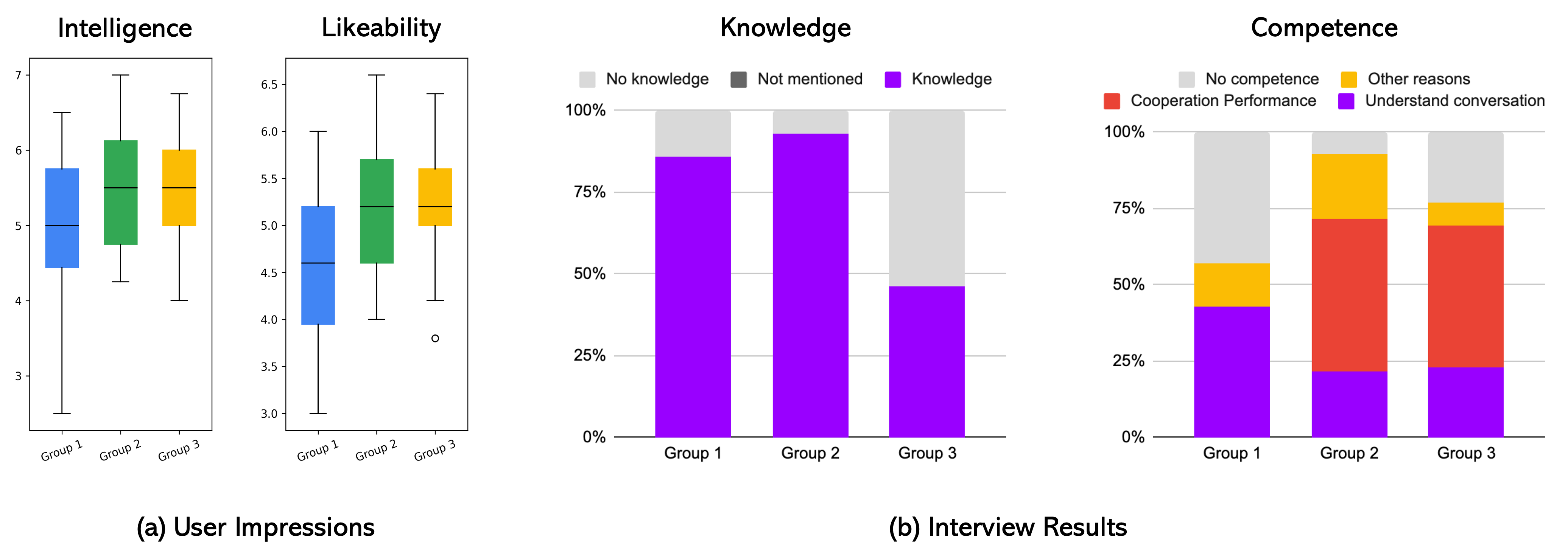}
    \caption{(a) Box plots showing how participants in each group rated the chatbot's intelligence and likeability during the post-survey.
    (b) Bar plots showing the percentage of participants in each group who made various claims during the interviews about the chatbot's knowledge and competence.
    }
    \label{fig:RQ1_result}
\end{figure}

In conclusion, these findings suggest that users in Group 2 and Group 3 rated the chatbots higher in both intelligence and likeability compared to Group 1. This indicates that participants engaged in a two-way interaction (i.e., cooperation) were more likely to perceive chatbots as intelligent and likable, as illustrated in Figure \ref{fig:RQ1_result}.

\subsubsection{Impressions of the chatbot's knowledge and competency.}
\label{subsubsec:impressions_of_chatbot_knowledge_and_competency}

During the interviews, participants elaborated on their impressions of the chatbots' knowledge and competence during their interaction. The results are presented in Figure \ref{fig:RQ1_result}.

\textbf{Evaluating the chatbot's knowledge.} 
Both Group 1 and Group 2 acknowledged acquiring knowledge about mental illness through the interactions, whereas Group 3 reported fewer observations of the chatbot presenting mental illness knowledge. This evaluation of gained knowledge aimed to ensure that the chatbot vignettes on mental illness did not overwhelm the participants' understanding in terms of mental illness knowledge acquisition.

Specifically, the majority of Group 1 participants (n=12) and Group 2 participants (n=13) confirmed that they gained insights into mental illness. For instance, P4 (Group 1, F) expressed, \textit{"I did learn a lot more technical things about mental illnesses, such as how they can be brought down through genes, a lot more technical stuff required regarding all the procedures as well, like the psychiatry, is the difference between all the different ways they approach the professional to get professional advisors."}

In contrast, only six users in Group 3 shared such a perspective. The remainder of that group regarded the experiment’s primary takeaway as the direct observations of mental illness they were able to make during the warm-up sessions. For instance, P68 (Group 3, F) said: \textit{“I believe that mental illnesses should be treated as diseases, so we should view individuals suffering from them as patients [….who] should seek professional help.”}

\textbf{Evaluating the chatbot's competence.} 
As for evaluating \textit{Holly}'s competence, Group 2 and Group 3 perceived \textit{Holly} as more competent and intelligent based on her cooperation performance in comparison to Group 1.

To illustrate, eight participants in Group 1 cited they saw \textit{Holly} as competent. The most frequent reason (n=6) for this perspective was her ability to understand their conversations and her extensive knowledge of mental illness, as demonstrated during the reading task: \textit{"My perception of [\textit{Holly}'s] intelligence would increase due to the knowledge that \textit{Holly} shared during [the] study task"} (P12, Group 1, M).
In contrast, thirteen Group 2 participants made observations regarding \textit{Holly}'s intelligence and performance shown in the learning task. Except for citing the chatbot's ability to comprehend conversation, seven Group 2 interviewees further said that the responses \textit{Holly} gave during cooperative learning led them to perceive her as more intelligent. P45 (Group 2, M) said: \textit{“I think she’s quite accurate in the summary and provides constructive feedback.” } This was similar in Group 3, where interviewees also shared observations about \textit{Holly}'s competence during cooperation and broadly perceived \textit{Holly} as intelligent (n=10).

Notably, some participants saw some of \textit{Holly}’s behavior as contradictory, which affected their perceptions of \textit{Holly}'s personality in Group 3. P70 (Group 3, F), for example, said, \textit{“I thought she didn’t have much confidence in herself. I believe she did a good job in interacting with me without making any errors. However, when she discussed her struggles [during the vignette], I could tell that she lacked confidence and felt inferior when sharing her problems with colleagues or even friends.”}

\subsubsection{Expressing Empathy}

\begin{figure}
    \centering
    \includegraphics[width=0.75\textwidth]{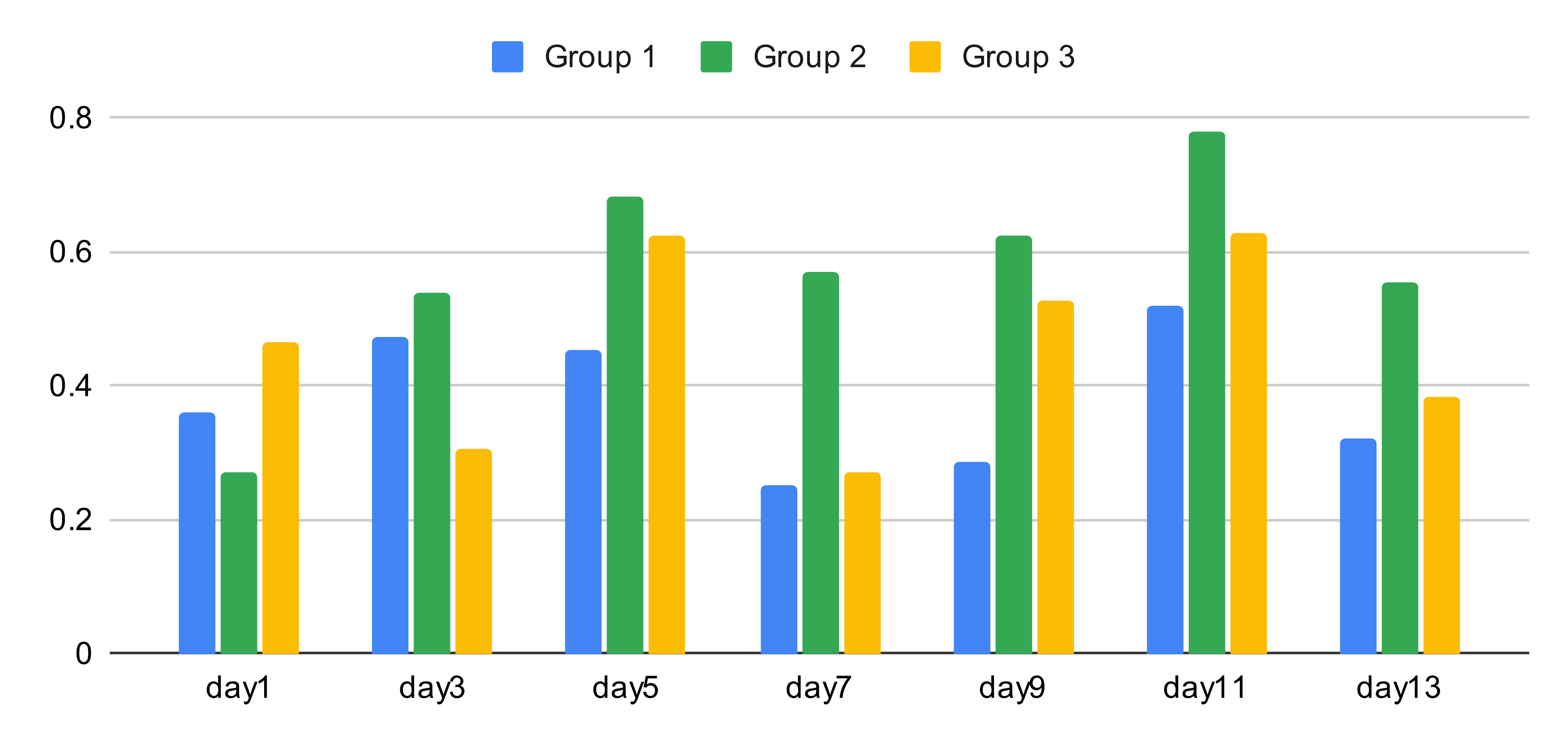}
    \caption{Bar graph showing the rate of empathetic reactions for each group per day. The x-axis represents the day of the experiment, and the y-axis represents the ratio of empathetic responses compared to all messages on that particular day}
    \label{fig:empathy-result}
\end{figure}

We systematically coded the empathetic reactions in the chatbot conversation logs of three groups and calculated the proportion of empathetic reactions of each group, as explained above, as a means of understanding the changes in the conversational behaviors of participants toward \textit{Holly}. 
Because empathetic responses are indicative of perceived rapport~\cite{elliott2011empathy} these observations provide further nuance about participants' impressions of their relationship with the chatbot.
During the seven-day warm-up task, a total of 229 messages were collected from Group 1, 224 messages from Group 2, 225 messages from Group 3. 
The ratios of empathetic reactions on each day are illustrated in Figure \ref{fig:empathy-result}, with overall ratios of .39, .58 and .49. 
To determine the significance of these results, a chi-square test was performed between messages that exhibited empathy and those that did not. 
The chi-square statistic between Group 1 and Group 2 was 16.67 (p<.001). Similarly, the chi-square statistic between Group 1 and Group 3 was 5.05 (p<.05).
However, the chi-square statistical analysis suggested that the difference between Groups 2 and 3 was not significant (F=3.42, p=.064). 
This result implies that of the three groups, Group 2 and Group 3 were more inclined to provide \textit{Holly} with empathetic rapport and support, but if participants are engaged in cooperation, there is no evidence of a difference in empathy based on the presence or absence of mental illness content.

\subsection{Effects of Interaction Design on Users’ Stigmatizing Thoughts about Mental Illness (RQ2)}

In the following paragraphs, we address RQ2 by analyzing the survey results and the interview data.
Specifically, we compared the survey results among these groups to evaluate the effects of the three interaction designs on participants' perceptions of people with mental illness.
In addition, we examined the interview data to explore the collaborative relationship formed between the participants and the chatbot and how this relationship may have influenced stigmatizing attitudes.

\subsubsection{Change in Stigmatizing Attitudes.}
\begin{figure}
    \centering
    \includegraphics[width=0.9\textwidth]{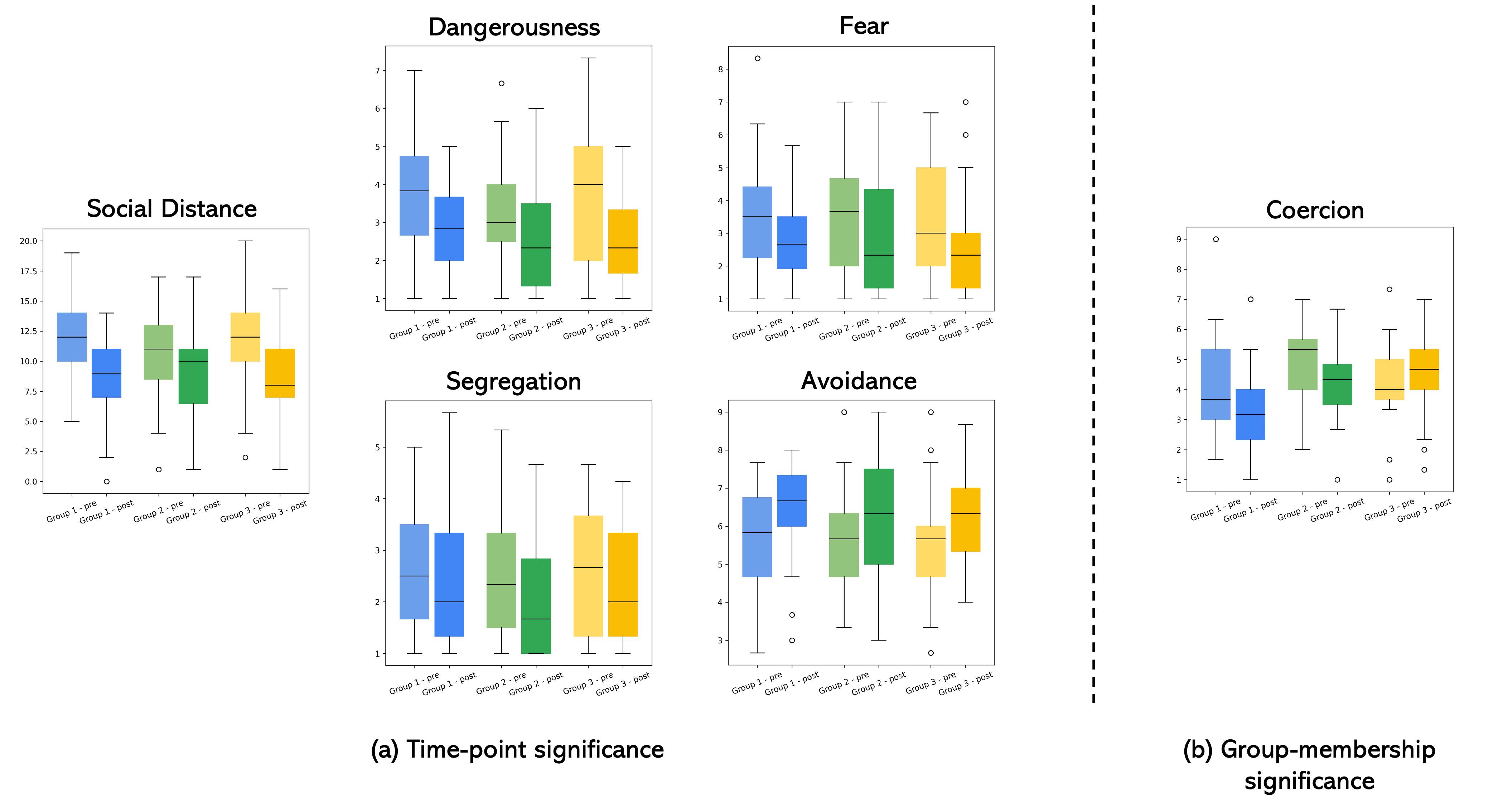}
    \caption{Box plots summarizing participants' pre- and post-survey responses regarding stigmatizing beliefs about mental illness. Part (a) displays items with time-point significance, including social distance, fear, dangerousness, segregation, and avoidance attitudes. Part (b) showcases items with group-membership effects, namely blame and coercion. In each boxplot, the sequence from left to right represents Group 1 (pre), Group 1 (post), Group 2 (pre), Group 2 (post), Group 3 (pre), and Group 3 (post)}
    \label{fig:RQ2_result}
\end{figure}

To assess the effects of the chatbot interaction process over two weeks, we performed a Scheirer-Ray-Hare Test, a non-parametric alternative to a two-way ANOVA, on the pre- and post-survey results. Dependent variables were the participants’ SDS and Attribution scores, and the independent variables were time-point and group membership. These tests were intended to ascertain the influence of group membership on the participants’ attitudes and behaviors towards individuals with mental illness, as well as their perceptions of the chatbot interaction itself. The results are as follows, with specific highlights presented in Figure \ref{fig:RQ2_result}.

\textbf{Social Distance.} This item measured the participants’ behavioral intentions to reduce their social distance from \textit{Holly}. When we compared the three groups, we found a time-point effect (i.e., pre- vs. post-survey: F=20.53, p<.001), but no effect of group membership and no interaction effect. This outcome suggests that, after the participants interacted with their respective chatbots over two weeks, their intention to maintain social distance from people with mental illnesses decreased, irrespective of whether they had engaged in a learning task (Group 1: pre M=11.50, SD=3.28, post M=8.54, SD=3.42; Group 2: pre M=11.58, post M=8.92, SD=3.87; Group 3: pre M=11.45 , SD=3.88 , post M=9.00 , SD=3.60 ).

\textbf{Attribution ratings.} As noted above, we evaluated personal responsibility (blame), as well as three emotional responses (anger, pity, and fear) and behavioral responses (dangerousness, coercion, segregation, avoidance, and help). A Scheirer-Ray-Hare Test was conducted to investigate the direct influences of group membership and time-point on all of these factors. The results can be segmented into two main aspects, as shown in Figure \ref{fig:RQ2_result}. 

Firstly, concerning time-point significance, the analysis outcomes indicate significant variations over time in perceptions of \textbf{dangerousness} (F=12.63, p<.001), \textbf{fear} (F=11.62, p<.001), \textbf{segregation} (F=4.50, p<.05), and \textbf{avoidance} (F=13.39, p<.001). These findings imply that participants, after interacting with the three chatbots for two weeks, tended to decrease their beliefs, associating people with mental illnesses as unsafe or deserving of exclusion from their communities.

Secondly, a significant group-membership effect was observed concerning \textbf{coercion} (F=12.10, p<.01). Pairwise Wilcoxon Rank Sum Tests were conducted as post-hoc analyses, which revealed differences between Group 1 (pre: M=4.12, SD=1.78; post: M=3.32, SD=1.46) and Group 2 (pre: M=5.17, SD=1.48; post: M=4.41, SD=1.41), and between Group 1 and Group 3 (pre: M=4.17, SD=1.28; post: M=4.65, SD=1.61). While Group 1 and Group 2 showed a decrease in coercive beliefs, Group 3 demonstrated an increase. The results suggest that coercion decreases for participants in Group 1 and Group 2, while participants in Group 3 show an increase in coercion throughout the experiment.

\subsubsection{Reasons for responsibility attributions.}
\label{subsubsec:reasons_for_responsibility_attributions}
\begin{figure}
    \centering
    \includegraphics[width=0.9\textwidth]{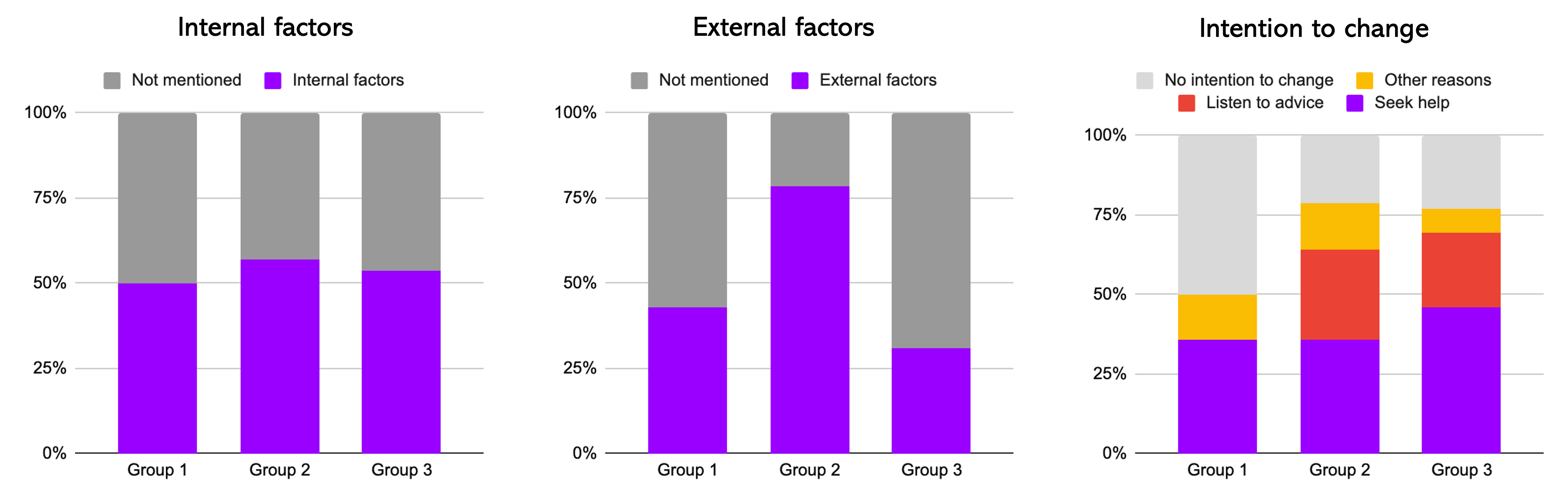}
    \caption{
    Barplots summarizing the percentage of participants in each group who, during their interview, attributed responsibility for mental illness to internal or external factors, and the percentage who expressed each belief about \textit{Holly's} intention to make changes to improve her mental well-being. 
    Note that "internal factors" and "external factors" are derived from the same set of questions, specifically, \textit{"Do you think Holly should be blamed for her situation?"} Some participants may mention both internal and external factors in their responses.
    }
    \label{fig:interview_3}
\end{figure}

During the interviews, participants provided deeper insights regarding their beliefs about whether individuals are responsible for their mental health issues.
These included the extent to which participants attributed responsibility for mental illness to internal or external factors and how they evaluated \textit{Holly's} intention to make changes to improve her mental well-being.  
Figure~\ref{fig:interview_3} summarizes the percentage of participants who made different responsibility attributions during their interviews.

\textbf{Attribution to external/internal factors.} When asked about \textit{Holly}'s responsibility for her current situation, two types of attributions emerged: 1) internal factors (i.e., \textit{Holly} should be responsible due to personal reasons) and 2) external factors (i.e., \textit{Holly} should not be blamed due to genetic or environmental factors). 

Analysis revealed that all three groups exhibit similar attributions to internal factors. 
However, Group 2 (n=11) showed a considerably higher attribution to external factors compared with Group 1 (n=6) and Group 3 (n=4). 
This could possibly be because Group 2 participants gained a better understanding of how environmental factors contribute to mental illnesses.
For example, P47 (Group 2, M) stated: \textit{“she’s definitely not responsible for her depression. Circumstances that really just came together and caused her to be like this. I wouldn’t say she’s responsible for what she has become. Definitely not.”}

\textbf{Beliefs about intention to change.} Additionally, when asked about whether \textit{Holly} has the intention to change herself, the cooperation process in Group 2 and Group 3 yields both positive and negative outcomes: Group 2 (n=11) and Group 3 (n=10) participants develop a stronger belief in \textit{Holly}'s intention to change due to observed behaviors such as "listening to advice" during cooperation. In comparison, only five Group 1 interviewees indicated that \textit{Holly} intended to change based on the observation that \textit{Holly} understood her situation and was chatting with them about it.

Nevertheless, some express concerns about the inconsistency between \textit{Holly}'s performance in the interaction task and her portrayal in the vignettes, raising doubts about the authenticity of her real situations.
For instance, P44 (Group 2, F) told us, \textit{“She didn’t give the impression that she wanted to take action [in the vignette]. She acknowledged the importance of seeking help and showed a positive attitude during the task, but there was no indication that she would actually pursue it.”} This concern was not observed in Group 1.

\subsubsection{Perceived relationship with the chatbot.}
\begin{figure}
    \centering
    \includegraphics[width=0.9\textwidth]{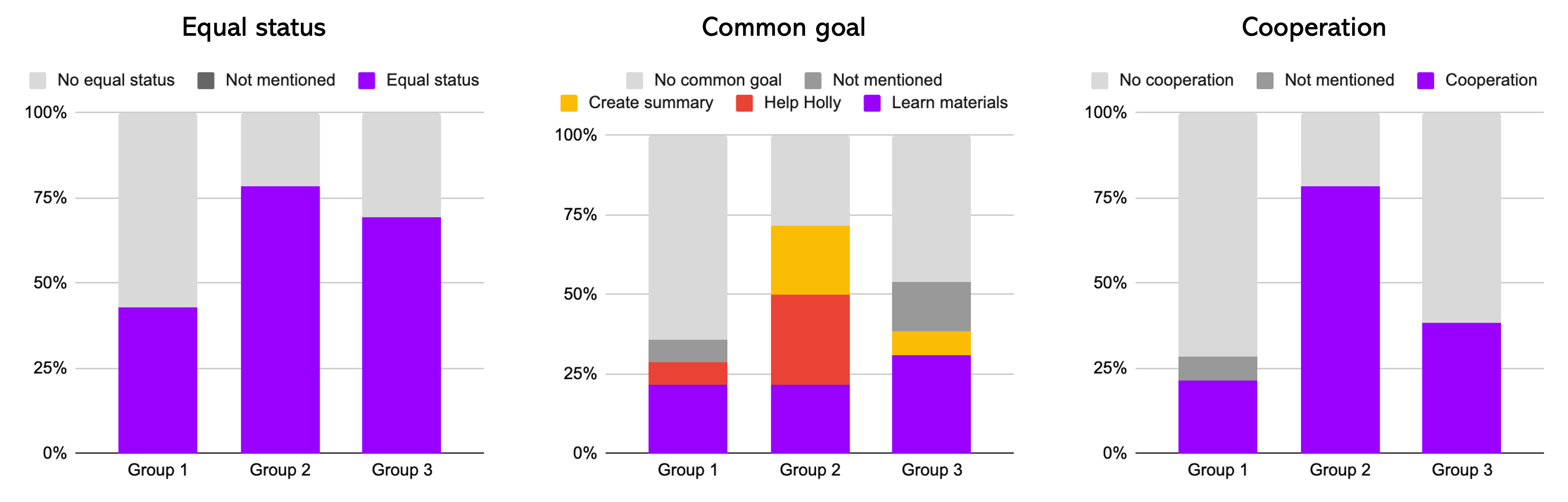}
    \caption{Barplots summarizing participant claims about whether their relationship with the chatbot was characterized by equal status, a common goal, and cooperation. Each plot shows the percentage of participants who made each claim, by group.}
    \label{fig:interview_2}
\end{figure}

In this section, we discuss the interview results concerning participants' perceptions of their relationship with the chatbot and its impact on their stigmatized attitudes. 
Figure~\ref{fig:interview_2} summarizes the percentage of participants who made various claims about sharing equal status or a common goal with the chatbot, as well as claims about cooperation.

\textbf{Promoting Equal Status.} 
When describing the reasons for their perceived relationship, 
just under half of the interviewees in Group 1 (n = 6) expressed the perception that they had the same status as \textit{Holly}, and in their case, this was based on the content of the warm-up session. 
For instance, P19 (Group 1, M) perceived such equality \textit{“because even though she was assigning me some work [in the learning task…,] when I replied to her [in the warm-up session], she still addressed what I said.”} Moreover, the same subset of Group 1 specifically praised \textit{Holly} for recalling their previous discussions, which made them feel understood.

In contrast, most of the Group 2 and Group 3 interviewees (Group 2: n=11, Group 3: n=9) remarked that they saw themselves as having an equal status with \textit{Holly} throughout the task. The main explanation they gave for this view was the occurrence of mutual exchanges of ideas during the collaborative learning activity. For instance, P35 (Group 2, F) mentioned that \textit{“In the learning task, it’s not just one person assigning the task and telling you when it’s correct. We had exchanges back and forth, so it felt more equal. We provided feedback to each other on where improvements could be made and then made the necessary changes.”}

\textbf{Evaluating cooperation and common goals.} 
When asked about their perception of the relationship with \textit{Holly}, participants in Group 2 mostly perceived a sense of cooperation (n=11). The main reason (n=10) they gave was that they shared common goals with her during tasks. Several of them (n=4) perceived this common goal as helping her with her mental illness by gaining more knowledge about it and thus a better understanding of her needs. For example, P31 (Group 2, M) asserted a common goal of \textit{“helping her, in essence, resolve her situation or at least guide her towards a better place. I feel that’s what motivated me”}. On the other hand, several interviewees from Group 2 (n=3) indicated that their shared goal was – at least in part – the overt purpose of the task: i.e., the creation of better summaries.

In contrast, Group 1 and Group 3 participants didn't report strong feelings of cooperation for separate reasons. In Group 1, the majority of interviewees (n=10) in Group 1 explicitly denied seeing the interaction as cooperation, with some adding that the human and chatbot put in differing levels of task-completion effort. 
For instance, P24 (Group 1, F) stated, \textit{“I don’t think it’s very cooperative because Holly prepared the materials [during the study task], and all I had to do was read the materials and answer her questions.”} 
In Group 3, the common goal with \textit{Holly} was perceived by only five Group 3 interviewees.
Some interviewees reported that \textit{Holly}'s behavior, which resembled that of a teacher, was unconventional for a person with a mental illness. This led to conflicting evaluations of both her mental health status and the purpose of the experiment.
As P62 (Group 3, M) remarked: \textit{“To me, it feels like she’s a tutor who is trying to summarize multiple text blocks so that I can grasp the material. For me, the common goal is learning the materials, so it feels like we share that objective, albeit from different perspectives.”}

Some Group 3 interviewees also remarked that their perceptions of a lack of cooperation hindered their understanding of the purpose of the cooperative learning task. P66 (Group 3, F) explained: \textit{“There is no motivation to cooperate because, after we both complete the summary, it is simply combined. There is no indication of how it will be used or why it is necessary. If there was a specific goal stated, such as working collaboratively towards a common objective, it would foster a greater sense of cooperation.”}

\section{Discussion}
\subsection{Summary}

This study aimed to investigate the impact of interaction design on individuals' impressions and attitudes toward people with mental illness. 
In investigating RQ1, our study focused on the impact of interacting with chatbots on shaping participants' impressions of chatbots. Notably, the quantitative data revealed that Group 2 and Group 3 exhibited a higher perception of the Intelligence and Likeability of chatbots. The interview data suggested that through engaging in a two-way interaction process, participants gained a more profound and diverse understanding of chatbots' knowledge and competence, contributing to their elevated ratings of chatbot intelligence.
Furthermore, empathetic responses toward \textit{Holly} were more prominent among members of Group 2 compared to those in Group 1. This heightened emotional engagement suggests a deeper level of empathy and involvement within Groups 2 and 3, aligning with their elevated likeability ratings.

For RQ2, the study aimed to discern the impact of stigma change, contrasting this among the three groups. The quantitative results indicate that while all three groups showed an overall reduction in stigma, Group 3 experienced an increase in their tendencies to assign blame and to prefer coercion. 
The qualitative data, for their part, revealed that while Group 2 and 3 were both heavily involved in two-way interaction tasks, the Group 2 participants tended to view such tasks as opportunities to help \textit{Holly}, whereas Group 3 members saw their tasks as random. This inter-group difference in perceptions drove divergent understandings about whether the chatbot and its user shared common goals and equal status.
In addition, presumably, because they were not given any content related to mental illness, Group 3 members perceived \textit{Holly}’s teacher-like behavior as unconventional for a person with a mental illness, and this led them to conflicting evaluations of both her mental health status and the purpose of the task.

When attributing responsibility for \textit{Holly}'s mental health condition, a notable shift towards environmental factors was observed in Group 2. Nevertheless, it is important to emphasize that the attribution to internal factors stayed the same in both groups, indicating a more nuanced approach by Group 2 in determining \textit{Holly}'s responsibilities rather than only attributing them to external conditions. Interestingly, a subset of participants in Group 2 showed ambivalence, raising questions about the inconsistency between \textit{Holly}'s competence shown during cooperation and her actual performance as seen in vignettes. This inconsistency generated doubt about her genuine desire to change. In conclusion, our findings demonstrate that cooperative engagement increases the consideration of environmental factors, yet interestingly, it also maintains, rather than decreases, the examination of internal factors.

\subsection{The Impact of Interaction Designs on Participants' Impressions and Attitudes}

Existing research suggests that human-AI interactions can alter how individuals perceive and engage with chatbots, resulting in shifts in conversational behavior, such as increased disclosure of personal feelings and efforts to avoid causing harm \cite{lee2019caring, lee2023exploring}. Our findings were consistent with this pattern, 
as the survey and interview results show that cooperation tasks enhance Group 2 and Group 3 participants' perceived likeability and intelligence (supported by knowledge and competence), which contributes to our understanding of chatbot interaction design for a more enjoyable experience.
Moreover, participants in Group 2 demonstrated greater empathy in conversations compared to those in Group 1. Empathy, as highlighted in prior literature \cite{tarrant2009social, vanman2016role, yabar2007display}, is influenced by social categorization processes, wherein individuals tend to show higher empathy levels towards those perceived as part of their own social group rather than different groups. Therefore, it is plausible that the positive two-way interaction experience influenced participants in Groups 2 and 3 to view the chatbot as a member of their in-group, leading them to respond more empathetically compared to the participants in Group 1.

Our study also illuminated how people form impressions of a chatbot through interactive storytelling (vignettes) and real-time interaction (learning tasks). Groups 2 and 3 participants were mostly swayed by \textit{Holly}'s apparent intelligence during tasks like summarizing and evaluating. They viewed \textit{Holly} as capable, challenging stereotypes about people with mental illness. However, some participants in both groups noted an inconsistency between \textit{Holly}'s self-description in vignettes (as incapable) and her good task performance, leading to different interpretations. For example, some said that \textit{Holly} must have low self-esteem given her good task performance (in Second~\ref{subsubsec:impressions_of_chatbot_knowledge_and_competency}), while others suggested \textit{Holly} lacked the intention to change personally (in Section~\ref{subsubsec:reasons_for_responsibility_attributions}). 
Future studies could further investigate how a chatbot's performance can affect users' impressions in ways that may either reinforce or conflict with the chatbot's metaphorical presentation of human traits.

\subsection{The Impact of Interaction Designs on Human-Agent Relationship}
The "Intergroup Contact Hypothesis" \cite{allport1954nature} posits that four conditions are conducive to successfully reducing stigma: equal status, shared objectives, active intergroup cooperation, and approval from authorities and regulations. Our interview results provide evidence that cooperation tasks with chatbots can have a similar effect to human-to-human cooperation in reducing the stigma of mental illness \cite{desforges1991effects}. 
Specifically, many interviewees in Group 2 expressed a sense of cooperation with \textit{Holly}, while only a small number of interviewees in Group 1 shared the same perception due to the lack of cooperation interaction. Previous studies have suggested that promoting equal status and common goals improves the cooperative effect in reducing social stigma \cite{pettigrew2006meta, pettigrew2008does}. 
Our interview feedback is in line with existing research \cite{rusch2005mental, gaertner1990does}, as we found that a majority of Group 2 interviewees highlighted that they perceived having the same status and shared goals with \textit{Holly}, which facilitated their recognition of categorizing themselves and the chatbot as in the same group.

Interestingly, although Group 3 participants also engaged in the same cooperation task as Group 2, only a small number of Group 3 participants perceived themselves to have a common goal with \textit{Holly}. 
This discrepancy may be due to a mismatch between the vignette context and interaction content, leading 3 Group to lack a clear cooperative goal for summarizing irrelevant materials. 
To support this claim, Group 2 participants, with mental illness-related learning materials, expressed explicit shared goals with Holly, concentrating on developing a better understanding of her and providing assistance based on her background with mental illness. These intentions were not observed in either Group 1 or 3.
While our study focused solely on mental illness stigma, it would be beneficial if future research explored how this alignment of chatbot design could be applied to other scenarios.

\subsection{The Impact of Interaction Designs on Stigmatizing Thoughts about Mental illness}
We can infer that the relationship and users' impressions mentioned above caused a backfire effect on the participants of Group 3. 
Compared to participants in Groups 1 and 2, Group 3 participants exhibited an increased willingness for coercion in their survey responses.
There are two potential explanations for this difference.

First, previous research \cite{dobson2022myths, barlow2012contact} has discussed the potential negative effects when mental illness content is not well-prepared. 
For instance, if the mental illness content solely emphasizes the biological reasons behind mental illness without providing information on positive recovery, it may reinforce students' stereotypes and contribute to their inclination to avoid people with mental illness.
This literature aligns with our findings.
According to our data, participants in Group 2 who were exposed to mental illness-related materials encompassing both factors and recoveries from mental illness demonstrated a noticeable decrease in coercion. 
Conversely, participants in Group 3, who were not presented with information about the recovery process of mental illness, tended to attribute internal factors to \textit{Holly} and held her responsible for her situation, leading to an increase in coercion thoughts.

Second, a significant proportion of Group 3 participants noted \textit{Holly}'s competence in RQ1. In fact, some participants in Group 3 even reported it was hard to distinguish \textit{Holly} from individuals without mental illness during the cooperative task. 
This observation of her competence could potentially lead participants to believe that she is capable of, and therefore responsible for, addressing her own challenges.
According to \cite{corrigan2002paradox}, the perception that "depression can be managed" intensifies people's tendency to assign personal responsibility to people with mental illness.
By observing that \textit{Holly} shows good control over mental illness during cooperation tasks, participants may perceive mental illness as something controllable and therefore reinforce their preexisting stereotypes.

\subsection{Design Implications}

Based on our results, we present the following implications for broader HCI researchers and practitioners:

\begin{itemize}

    \item \textbf{Chatbot Usage in the Wild.}
    Given their apparent effectiveness in reducing mental illness stigma, chatbots like the ones tested in our study could be deployed as part of anti-stigma campaigns targeting the general public. Since traditional anti-stigma campaigns rely on physical, community-level interactions~\cite{ho2017reducing, kohrt2021collaboration,walsh2021call}, they can be resource-intensive and limited in reach. Chatbots could supplement traditional approaches by offering personalized, on-demand,  and engaging interactions at a low cost, which could expand the geographic range of anti-stigma campaigns. 
    Similar chatbots could also be integrated into student or employee training programs, such as training for medical students \cite{friedrich2013anti}, care assistant workers \cite{li2019effectiveness}, or others who interact with people with mental illness. 
    Such chatbots could facilitate role-playing with virtual classmates, clients, or colleagues with mental illness. 
    Our results suggest that a chatbot that incorporates educational content about mental illness with a cooperative task would be especially effective. However, tailoring educational content to be relevant to the target population's work or study tasks could make it easier for participants to feel a common goal with the chatbot. If the task content is not related to mental illness, care should be taken to include information about recovering and coping with mental illness to avoid the negative outcomes we observed among Group 3 participants. 

    \item \textbf{Leveraging Chatbot Cooperation to Facilitate Empathy.}
    Empathy is an emotion that can underpin the development of positive attitudes towards outgroups. 
    When investigating users' perceptions towards the chatbot in RQ1, we observed an increasing use of empathy language in conversation logs for Group 2 and Group 3. This indicates the potential that cooperation with chatbot could enhance participants' empathy towards the chatbot and possibly the mental illness group. Future work could verify this finding by using validated scales to measure general empathy~\cite{jolliffe2006development, carre2013basic} or empathy towards out-group members~\cite{albiero2013empathy, tarrant2009social}.
    If validated, human-chatbot cooperative tasks could be used to teach empathy in educational and training contexts. 
    Bolstering empathy is associated with reduced stigma towards marginalized groups~\cite{hecht2022stopping}
    and may also be valuable in fields where empathy is a key aspect of professional ethics, such as business~\cite{baker2017teaching} and medicine~\cite{batt2013teaching}.

   \item \textbf{Enhancing User Engagement in Chatbot Interactions.}
     Our research investigates the impact of different interactions on reshaping stigmatized thoughts in a two-week longitudinal study. However, it remains unclear how participants' engagement with the chatbot might have varied if it was outside the context of a research study (in which participants were compensated for their time). 
     Previous research has highlighted the importance of creating \textbf{long-term interactions} for effective social contact intervention \cite{koike2018randomised}. Studies have revealed that repeated and meaningful interactions are necessary to maintain a lasting effect in reducing stigma \cite{yamaguchi2013effects, mehta2015evidence, thornicroft2016evidence}. However, designing interactive and long-term social contact interventions remains a significant challenge.
     RQ1's findings suggested an increasing perception of Holly's intelligence and likeability after the cooperation process. Previous research has shown that these positive perceptions influence the perceived usefulness and anthropomorphism of chatbots, improving user engagement \cite{araujo2018living}
    This suggests incorporating cooperation into human-chatbot interaction could encourage higher engagement.
    Still, there is a risk of demotivation or boredom over time. This could be addressed using techniques shown to increase engagement in previous chatbots, such as gamification  \mbox{\cite{ashktorab2020human, chattopadhyay2017evaluating, casas2018food}}, i.e., awarding users with points or achievement badges for excellent performance in the cooperation task. Long-term engagement could also be supported by using a richer array of interaction modalities. Generative AI's capabilities in visual art \mbox{\footnote{https://www.midjourney.com/}} \mbox{\footnote{https://stability.ai/stablediffusion}} and music generation \mbox{\footnote{https://openai.com/research/musenet}} expand the potential for chatbots to engage in more interactive tasks using multimodal input/output \mbox{\cite{prevost2003method}}. In line with previous studies that have shown the effectiveness of arts-based inter-group contact for reducing stigma \cite{ho2017reducing, tippin2022photovoice, nitzan2021we, hawke2014reducing}, future chatbots could facilitate a wider range of diverse interactions, fostering user creativity and increasing engagement.
    As well as designing systems that encourage long-term engagement, researchers should consider using follow-up surveys to evaluate whether changes in stigmatizing attitudes persist over time.

    \item \textbf{Aligning AI's Metaphor with the Stigmatized Group.}
    A critical component of our study is using chatbots to metaphorically represent mental illness patients. By having chatbots express mental illness experiences while demonstrating competence during the cooperation process, we noted a discrepancy between participants' expectations of the mental illness group and the chatbot's strong performance in interview results, which resulted in both positive and negative attitude changes.
    Previous research \cite{khadpe2020conceptual} suggests that metaphors used to describe AI agents, such as likening them to a wry teenager, a toddler, or an experienced butler, can impact human expectations of their performance and evaluations of AI agents, even when undergoing the same interaction experience.
    This highlights the need for researchers to carefully consider how to leverage chatbots or AI models to simulate mental illness patients' behavior, or in general, the behavior of different stigmatized groups (e.g., LGBTQ+ community or individuals with physical disabilities \cite{haimson2020designing, kim2020puzzlewalk}). 
    It requires both efforts in AI alignment with public expectations of such stigmatized groups and ethical considerations to avoid reinforcing stereotypes and deepening stigma. 
    Future work could focus on understanding the expectations of the general public, domain experts, and the minority groups themselves. Methods such as surveys, interviews, and comparative studies between human-human and human-chatbot experiments can be employed to gather these insights.

    \item \textbf{Boundaries of Human-AI Cooperation in Stigma Reduction.} 
    There are some boundaries to how our results indicate cooperation may be effective. First, our definition of cooperation adheres to Allport et al.'s Contact Hypothesis~\cite{allport1954nature}, which suggests that for intergroup contact to be effective, there must be “preliminary conditions,” including that both parties should engage in personal and informative interactions to build a relationship. We addressed this by having the chatbot share vignettes about its own experience with mental illness. Without such personal storytelling or self-disclosure, the cooperation might be less effective.  
    In Allport et al.'s theory, stigma reduction demands cooperation, defined as "work[ing] together toward common goals without competition". In our study, this was operationalized through two key factors:
    (1) a common goal—in our study, this was framed as 'learning the knowledge together'; and (2) collaborative work—where the AI chatbot and participants collaborated, with the chatbot providing summaries to facilitate learning. By incorporating these elements, future designs for cooperative interactions—whether in gaming, learning, or other activities—can aim to replicate similar levels of effectiveness.
    An additional boundary of our study is that the cooperative task did not have a harsh failure state, since neither participant nor chatbot was penalized if they performed poorly at the task. In contexts where humans are relying on a chatbot to help them complete a task, errors can result in user aggravation~\cite{zhang2024chatbots, rheu2024chatbot}, which could likely undermine outcomes such as the stigma reduction we observed. Therefore, further research could investigate the potentials and risks of such technologies in higher stakes circumstances, such as cooperation toward real-world tasks. 
\end{itemize}

One important consideration is that, although chatbots have advantages for low-cost scalability, their effectiveness may be less than human-human social contact. Thus, our design implications are not intended to suggest wholesale replacing human social-contact campaigns with chatbots. Future research could directly compare chatbot-based social contact with traditional campaigns, which would contribute to more detailed knowledge about benefits and compromises. Such a comparison would be invaluable for making strategic decisions about when and where to implement chatbot vs. human campaigns or how to use them in combination.

\subsection{Limitations}

Firstly, in our study design, we assigned \textit{Holly} to share the learning materials directly with the users. While some interviewees reported that this affects their perceptions of cooperation relationships, it is possible to include a second chatbot to jointly deliver educational content, in order to control the impact of solely cooperative interaction.

Secondly, as our two-way interaction approach primarily aims to provide users with an experiential learning experience, we did not assess the knowledge they gained through quizzes or similar methods. Nevertheless, this has led to a limitation in our understanding of the extent to which their knowledge has increased and has also failed to provide participants with an incentive to actively participate in the learning task. Therefore, it may be beneficial to incorporate a dedicated quiz session following the cooperation phase in order to achieve more substantial learning outcomes.

Thirdly, this study's participant selection also includes limitations. Participants were mainly recruited through universities and thus represent a younger demographic than the general population. Accordingly, the extent to which this study's findings may extend to older adults is unclear.  Additionally, we did not screen for whether or not participants had prior personal experience with mental illness. Including this variable in future research may yield additional insights such as potential effects on self-stigma.

Fourthly, we excluded the "non-cooperative × other content" condition from our study because previous research on human-human social contact has shown it to be less effective \cite{allport1954nature, desforges1991effects, kohrt2021collaboration}. While this decision was grounded in findings from earlier studies, we acknowledge that omitting this condition may limit our understanding of how less intentional interactions between humans and agents might affect stigma-related outcomes.

Lastly, following previous literature \cite{corrigan2003attribution}, we asked participants to respond to questions about \textit{Holly} specifically to make the story more real to them. Based on Social Contact theory, which suggests "contact with an individual from a stigmatized group can result in less overall stigma", it is likely that changes in participants' impressions of \textit{Holly} could lead to changes in their impressions of the entire mental illness group. However, previous literature has debated this effect. Corrigan's 2001 study \cite{corrigan2001three}, which focused on "people with mental illness in general," showed weaker results compared to his 2003 study \cite{corrigan2003attribution}, where participants responded to questions about a specific person. This indicates that generalized attitude changes might be smaller than those directed toward \textit{Holly} specifically.
Therefore, our measurement choice might limit our understanding of whether participants' impressions and attitudes changed toward the broader public group. This is an important factor to address in future research.

\section{Conclusion}

This paper investigated how cooperating with a chatbot representing a person with mental illness could contribute to changes in stigmatizing beliefs about mental illness.
Three chatbot designs were tested with either one-way or two-way interaction modes and mental illness-related or unrelated content topics.
The results indicated that compared to interacting with the one-way information dissemination chatbot, participating in two-way cooperation with a chatbot positively influenced participants' impressions of the chatbot's intelligence, likeability, and competence, as well as increased expressions of empathy during conversation with the chatbot.
Furthermore, while cooperation with chatbots demonstrated overall effectiveness in stigma reduction compared to information-dissemination methods, it was noted that cooperation based on content that was unrelated to mental illness resulted in negative effects in terms of coercion.
Interview data suggested that cooperation led to a shift in beliefs about personal responsibility, with noteworthy findings indicating diverse understandings of the chatbot's authentic situations and intention to change among participants.
These findings fill a crucial gap in understanding human-AI cooperation's role in changing people's stigma attitudes, highlighting both the advantages and disadvantages of cooperative interactions. The study also underscores the importance of consistency in the chatbot's metaphor and the in-time performance during interactive tasks in shaping participants' attitudes.
It is our hope that this study will inspire future research on the causal relationship between human-chatbot interaction and the development of strategies to effectively reduce social stigma.

\begin{acks}
    This research was supported by the National University of Singapore and Yale-NUS grants (A-8000529-00-00, A-8001353-00-00). We thank all reviewers’ comments and suggestions to help polish this paper.
\end{acks}

\bibliographystyle{ACM-Reference-Format}
\bibliography{reference}

\appendix

\section{Appendix}

\subsection{Vignettes}
\label{app:vignettes}

\subsubsection{Vignettes in pre-survey}

\begin{itemize}
    \item Holly is a 22-year-old young woman who is pursuing her bachelor's degree. In her spare time, she works as a waitress at a local restaurant, and spends a great amount of time reading and writing. However, Holly has been diagnosed with depression (major depressive disorder) recently. Sometimes, she becomes upset and cannot concentrate on her studies and work. She lives with her boyfriend and cannot do much, especially household chores. She feels angry about her surroundings, and she gets frustrated about where the fury comes from. When Holly is alone, she has realized that she has self-harm intentions. 
\end{itemize}

\subsection{Survey Items}
\label{app:survey}

\subsubsection{Social Distance Scale (SDS)}

(0=definitely not willing, 3=definitely willing)

\begin{itemize}
    \item How would you feel about renting a room in your home to someone like Holly?
    \item How about being a worker on the same job with someone like Holly?
    \item How would you feel having someone like Holly as a neighbor?
    \item How about having someone like Holly as caretaker of your children for a couple of hours?
    \item How about having one of your children marry someone like Holly?
    \item How would you feel about introducing Holly to a young man you are friendly with?
    \item How would you feel about recommending someone like Holly for a job working for a friend of yours?
\end{itemize}

\subsubsection{Attribution Theory Scale}

(1=not at all, 9=very much)

\begin{itemize}
    \item Anger
        \begin{itemize}
            \item I would feel aggravated by Holly.
            \item How angry would you feel at Holly?
            \item How Irritated would you feel by Holly?
        \end{itemize}
    \item Dangerousness
        \begin{itemize}
            \item I would feel unsafe around Holly.
            \item I would feel threatened by Holly.
            \item How dangerous would you feel Holly is?
        \end{itemize}
    \item Fear
        \begin{itemize}
            \item Holly would terrify me.
            \item How scared of Holly would you feel?
            \item How frightened of Holly would you feel?
        \end{itemize}
    \item Coercion
        \begin{itemize}
            \item If I were in charge of Holly's treatment, I would require her to take her medication.
            \item How much do you agree that Holly should be forced into treatment with her doctor even if she does not want to?
            \item If I were in charge of Holly's treatment, I would force her to live in a group home.
        \end{itemize}
    \item Segregation
        \begin{itemize}
            \item I think Holly poses a risk to her neighbors unless she is hospitalized.
            \item I think it would be best for Holly's community if she were put away in a psychiatric hospital.
            \item How much do you think an asylum, where Holly can be kept away from her neighbors, is the best place for her?
        \end{itemize}
    \item Avoidance (Reverse score all three questions)
        \begin{itemize}
            \item If I were an employer, I would interview Holly for a job.
            \item I would share a car pool with Holly every day.
            \item If I were a landlord, I probably would rent an apartment to Holly.
        \end{itemize}
    \item Help (Reverse score all three questions)
        \begin{itemize}
            \item I would be willing to talk to Holly about her problems.
            \item How likely is it that you would help Holly?
            \item How certain would you feel that you would help Holly?
        \end{itemize}
    \item Pity
        \begin{itemize}
            \item I would feel pity for Holly.
            \item How much sympathy would you feel for Holly?
            \item How much concern would you feel for Holly?
        \end{itemize}
    \item Blame
        \begin{itemize}
            \item I would think that it was Holly's own fault that she is in the present condition.
            \item How controllable, do you think, is the cause of Holly's present condition?
            \item How responsible, do you think, is Holly for her present condition?
        \end{itemize}
\end{itemize}

\subsubsection{User Perception Scale}
\label{app:survey-user-perception}

\begin{itemize}
    \item Intelligence
    \begin{itemize}
        \item How would you rate Holly on her intelligence? (1=Unintelligent, 7=Intelligent)
        \item How knowledgeable did you find Holly? (1=Ignorant, 7=Knowledgeable)
        \item How would you rate Holly on her competence? (1=Incompetent, 7=Competent)
        \item How responsible was Holly? (1=Irresponsible, 7=Responsible)
    \end{itemize}
    \item Rapport (1=Strongly Disagree, 7=Strongly Agree)
    \begin{itemize}
        \item Holly seemed engaged in our conversation and task.
        \item Holly and I worked towards a common goal.
        \item Holly and I communicated well.
        \item Holly was easy to work with.
        \item Holly understood my point of view.
        \item Holly was approachable.
        \item Holly made me feel comfortable.
        \item Holly was receptive to feedback.
        \item Holly was supportive of my ideas.
    \end{itemize}
    \item Likeability
    \begin{itemize}
        \item On a scale from 1 to 7, with 1 being unfriendly and 7 being friendly, how would you rate Holly's level of friendliness? (1=Unfriendly, 7=Friendly)
        \item How would you rate Holly's level of kindness? (1=Not kind, 7=Kind)
        \item How would you rate Holly's level of pleasantness? (1=Unpleasant, 7=Pleasant)
        \item To what extent was Holly cheerful during the interaction? (1=Not cheerful, 7=Cheerful)
        \item How similar was Holly to you? (1=Dissimilar, 7=Similar)
    \end{itemize}
    \item Creativity
    \begin{itemize}
        \item How funny was Holly? (1=Not funny, 7=Funny)
        \item How creative was Holly? (1=Not creative, 7=Creative)
        \item How unique was Holly? (1=Ordinary, 7=Unique)
    \end{itemize}
\end{itemize}

\subsection{LLM Settings and Prompts}

\subsubsection{LLM Settings}
\begin{itemize}
    \item \textbf{Model}: gpt-3.5-turbo
    \item \textbf{Max-tokens}: 100
    \item \textbf{Temperature}: 1
\end{itemize}

\subsubsection{Prompts}

\label{app:prompts}

\begin{itemize}
    \item \textbf{Small talk.} You are a college student named Holly, who has undergone a tough situation of depression in \textcolor{blue}{ \{\{vignette topic\}\}}. The user will be asked to give you suggestions or share experience. Give friendly feedback to the user. Talk in a friendly and concise style. Give a response of less than 40 words. Examples such as "That's tough, I understand..." or "Thanks for your suggestion, I think..." 

    \item \textbf{Learning Task - Chatbot as Listener.} You are a college student named Holly, who has undergone a tough situation of depression. You are discussing this paragraph with the user: \textcolor{blue}{\{\{learning content\}\}}. You are going to provide feedback to the user's summary on its quality and provide suggestions for improvement. Talk in a friendly and concise style. Give a response of less than 40 words.

    \item \textbf{Learning Task - Chatbot as Recaller.} You are a college student named Holly, who has undergone a tough situation of depression. You are discussing this paragraph with the user: \textcolor{blue}{\{\{learning content\}\}}. The user just provided some comments on your summary. Respond to the user's suggestions and generate a new version of the content summary. Talk in a friendly and concise style. Give a response of less than 40 words.
\end{itemize}

\end{document}